\documentclass[a4paper,11pt]{article}
\pdfoutput=1
\usepackage{jheppub}
\usepackage[T1]{fontenc}
\usepackage{booktabs}
\usepackage{array}
\newcolumntype{L}[1]{>{\raggedright\arraybackslash}p{#1}}
\usepackage{longtable}
\usepackage{pdflscape}
\usepackage{enumitem}
\setlist{nosep,leftmargin=*}
\newcommand{\TeV}{\,\mathrm{TeV}}
\newcommand{\GeV}{\,\mathrm{GeV}}

\newcommand{\met}{p_{T}^{\mathrm{miss}}}
\newcommand{\gagg}{g_{a\gamma\gamma}}
\newcommand{\gann}{g_{a\nu\nu}}
\newcommand{\sqrts}{\sqrt{s}}
\newcommand{\Zsys}{Z_{1\%}}
\title{\boldmath  Axion-Like Particle Searches at Muon Colliders: Photon–Neutrino Signatures and Heavy-Neutral-Lepton Feasibility}

\author[a,*]{Ijaz Ahmed}
\author[b]{Waqas Ahmad}
\author[a]{M. Tayyab Javaid}
\author[c]{Jamil Muhammad}
\affiliation[a]{Federal Urdu University of Arts, Science and Technology,
Islamabad, Pakistan}
\affiliation[b]{Riphah International University, Islamabad, Pakistan}
\affiliation[c]{Sang-Ho College \& Department of Physics, Konkuk University, Seoul 05029, South Korea}
\note[*]{Corresponding author.}
\emailAdd{ijaz.ahmed@fuuast.edu.pk}

\abstract{We study the mono-photon signature
$\mu^+\mu^-\to\gamma a\to\gamma\nu\bar\nu$ at $3$ and $10\TeV$ muon colliders,
with integrated luminosities of $1$ and $10\,\mathrm{ab}^{-1}$, in a
two-coupling axion-like-particle (ALP) simplified model. Five benchmarks
spanning $m_a=100$--$1000\GeV$ are generated at
$g_{a\gamma\gamma}=0.01\TeV^{-1}$ and flavor-universal pseudoscalar neutrino
couplings $y_\nu=0.1$ or $0.3$, giving invisible branching fractions above
99.995\% and mono-photon cross sections of $0.00854$--$0.01217\,\mathrm{fb}$.
Assuming a 1\% background normalization uncertainty, a cut-based photon-line
selection gives projected 95\% sensitivity to
$g_{a\gamma\gamma}\simeq0.039$--$0.045\TeV^{-1}$ at $3\TeV$ and
$0.0137$--$0.0140\TeV^{-1}$ at $10\TeV$. A ROOT/TMVA gradient-boosted decision
tree with validation-fixed thresholds reaches held-out AUC values of
$0.9967$--$0.9993$ and $g_{95}=0.0192$--$0.0400\TeV^{-1}$ at $3\TeV$ and
$0.0122$--$0.0144\TeV^{-1}$ at $10\TeV$. Because only 2--41
irreducible-background test events populate the optimized classifier tails,
the mass--coupling contours are reported as projected expected sensitivities
rather than profile-likelihood or CL$_s$ exclusions. A separate appendix records a phenomenological, parton-level HNL diphoton feasibility study; it is not used to claim detector-level sensitivity or exclusion.}

\begin{document}
\maketitle
\flushbottom

\section{Introduction}
\label{sec:introduction}

Axion-like particles are generic pseudo-Nambu--Goldstone bosons whose
low-energy interactions with Standard Model fields are naturally organized in
an effective-field-theory expansion~\cite{Jaeckel:2010ni,Brivio:2017ije,
Bauer:2020jbp}. Their photon coupling motivates associated production at
high-energy lepton colliders, while invisible ALP decays lead to a clean
mono-photon topology~\cite{Bauer:2017ris,Bauer:2018uxu}. Multi-TeV muon
colliders provide a particularly useful environment because they combine a
clean initial state with high collision energy and large target integrated
luminosity~\cite{deBlas:2022aow,Accettura:2023}. Existing collider
constraints on the ALP--photon coupling come from LEP and LHC data for light
ALPs~\cite{Jaeckel:2015jla} and, for $m_a=500$--$2000\GeV$, from central
exclusive diphoton production with tagged protons at the LHC~\cite{CMS:2024ppsalp}.

This work studies $\mu^+\mu^-\to\gamma a$ followed by
$a\to\nu\bar\nu$ in a deliberately minimal two-coupling simplified model. We
first establish the model interpretation and its limitations, then document
the event generation, selections, cut-based results and multivariate analysis.
All benchmarks are listed in table~\ref{tab:benchmarks}; the two distinct
cut-based and TMVA preselections are given in table~\ref{tab:selections}. The
paper concludes with projected coupling sensitivities and a precise statement
of the assumptions needed to interpret them.

The closest previous mono-photon study considered the same 3 and 10 TeV
staging scenarios, but treated an invisible light ALP or dark photon rather
than an ALP decaying through an explicit neutrino interaction
~\cite{Casarsa:2021rud}. Electroweakly coupled ALPs at muon colliders have
also been studied in vector-boson-fusion and associated production channels
~\cite{Bao:2022onq}, while recent lepton-collider work emphasizes
complementary long-lived and off-shell regimes~\cite{Bao:2025alp}.
Heavy neutral leptons are themselves a well-motivated target~\cite{Abdullahi:2023},
and muon colliders offer a strong search reach for them above the $Z$
pole~\cite{Mekala:2023,Kwok:2023}.
The ALP--HNL interplay at a future muon collider has also been studied in a
broader effective-field-theory framework, including HNL-pair production and
cascade signatures~\cite{BurgosMarcos:2025}. The present HNL analysis is
narrower and addresses the distinct $\nu_\mu\bar\nu_\mu a$ final state with
$a\to\gamma\gamma$, missing momentum, and a dedicated 3 and 10 TeV
parton-level selection; it is not the first study of ALP--HNL physics at a
muon collider. The paper's main contribution is the combined consistency
analysis of the ALP--neutrino mono-photon benchmark, its detector-level
cut-based and multivariate recoil study, and this separately scoped HNL
diphoton feasibility analysis. Table~\ref{tab:modelcomparison} distinguishes
these assumptions and non-comparable normalizations.

\section{Physics model, samples, and benchmark construction}

\subsection{Low-energy simplified Lagrangian}

The new state $a$ is a neutral, color-singlet, spin-zero pseudoscalar with mass
$m_a$. The operator normalization follows standard collider ALP
conventions~\cite{Brivio:2017ije,Bauer:2017ris,Bauer:2020jbp}. The collider
samples are defined by the phenomenological low-energy Lagrangian
\begin{equation}
 \mathcal L_{\rm simp}=
 \frac{1}{2}(\partial_\mu a)(\partial^\mu a)
 -\frac{1}{2}m_a^2a^2
 -i\sum_{i=e,\mu,\tau}y_{\nu_i}\,
 a\,\bar\nu_i\gamma_5\nu_i
 +\frac{\gagg}{4}aF_{\mu\nu}\widetilde F^{\mu\nu},
 \label{eq:lagrangian}
\end{equation}
where
\begin{equation}
 \widetilde F^{\mu\nu}\equiv
 \frac{1}{2}\epsilon^{\mu\nu\rho\sigma}F_{\rho\sigma}.
\end{equation}
The photon coupling $\gagg$ has mass dimension $-1$, while the neutrino
couplings $y_{\nu_i}$ are dimensionless. Flavor-universal neutrino couplings, $y_{\nu_i}=y_\nu$, are assumed. Both
$a\bar\nu i\gamma_5\nu$ and $aF\widetilde F$ are CP even when $a$ is CP odd,
so Eq.~\eqref{eq:lagrangian} preserves CP for real couplings. The interaction
is specified after electroweak symmetry breaking and should be regarded as a
simplified phenomenological description rather than a complete ultraviolet
theory.

For input-card convenience, the UFO introduces a dimensionful parameter
$\gann$ and defines
\begin{equation}
 y_\nu=m_a\gann,
 \qquad \gann=\frac{y_\nu}{m_a}.
 \label{eq:ynumapping}
\end{equation}
Thus $\gann$ is not an additional independent interaction in this scan: the
physical three-point pseudoscalar coupling is the dimensionless $y_\nu$. This
distinction is important when comparing benchmarks at different masses.

\subsection{UFO realization and operator isolation}
\label{sec:ufo}

In the coupling-clean UFO, constructed in the FeynRules/UFO
framework~\cite{Alloul:2013bka,Degrande:2011ua}, the ALP--neutrino vertex is
implemented for each flavor as
\begin{equation}
 \bar\nu_i\nu_i a:\qquad
 \texttt{GC\_80}=m_a\gann=y_\nu,
 \qquad
 \texttt{FFV\_ANu}=P_L-P_R=-\gamma_5,
\end{equation}
up to the overall UFO/Feynman-rule phase convention. The photon interaction is
implemented as
\begin{equation}
 \gamma\gamma a:\qquad
 \texttt{GC\_81}=\frac{i\gagg}{2},
\end{equation}
where the UFO Lorentz tensor contains twice
$\epsilon^{\mu\nu\rho\sigma}p_{1\rho}p_{2\sigma}$; together these factors
reproduce the normalization $(\gagg/4)aF\widetilde F$ in
Eq.~\eqref{eq:lagrangian}.

The generated cards set the remaining ALP coefficients to
\begin{equation}
 C_{\widetilde G}=C_{\widetilde W}=C_{\widetilde B}=C_{a\Phi}=0.
\end{equation}
Consequently, no gluonic, $aZZ$, $aZ\gamma$, $aWW$, or direct
ALP--charged-fermion operator contributes to the advertised signal. Only
$\gagg$ controls production and $y_\nu$ controls the invisible decay.

\subsection{Relation to derivative ALP--neutrino interactions}

Equation~\eqref{eq:lagrangian} must not be identified with a diagonal
derivative coupling to an ordinary massless active-neutrino current. For a
vector current,
\begin{equation}
 \frac{c_V}{f_a}(\partial_\mu a)\bar\nu\gamma^\mu\nu
 =-\frac{c_V}{f_a}a\,\partial_\mu(\bar\nu\gamma^\mu\nu)
 \longrightarrow 0,
\end{equation}
because the diagonal Dirac vector current is conserved. For an axial current,
the equations of motion give
\begin{equation}
 \partial_\mu(\bar\nu\gamma^\mu\gamma_5\nu)
 =2im_\nu\bar\nu\gamma_5\nu+\text{anomaly terms},
\end{equation}
so its on-shell pseudoscalar coupling to an ordinary active neutrino is
proportional to the tiny neutrino mass and vanishes in the massless limit.

The unsuppressed $y_\nu=0.1$--0.3 values used here therefore require a suitable
completion containing right-handed or sterile-neutrino degrees of freedom, or
must simply be interpreted as low-energy phenomenological parameters. The
present UFO uses Dirac-style $\nu_i\bar\nu_i$ final states and a
flavor-universal coupling. It is not, by itself, an
$SU(2)_L\times U(1)_Y$-gauge-invariant active-neutrino EFT. A UV-complete
sterile/heavy-neutrino model would introduce additional masses, mixing angles,
production vertices, and possibly visible decay modes; those ingredients are
outside the scope of the current samples. Explicit ALP--sterile-neutrino
completions illustrate these additional ingredients~\cite{Alves:2019xpc}.

\subsection{ALP decay widths and branching fractions}

For one Dirac neutrino flavor with mass $m_{\nu_i}$, the pseudoscalar decay
width following from Eq.~\eqref{eq:lagrangian} is
\begin{equation}
 \Gamma(a\to\nu_i\bar\nu_i)
 =\frac{y_{\nu_i}^2m_a}{8\pi}
 \sqrt{1-\frac{4m_{\nu_i}^2}{m_a^2}}.
 \label{eq:singleflavorwidth}
\end{equation}
The simulations take $m_{\nu_i}=0$ and three universal flavors, giving
\begin{equation}
 \Gamma_{\nu\bar\nu}
 =\sum_i\Gamma(a\to\nu_i\bar\nu_i)
 =\frac{3y_\nu^2m_a}{8\pi}
 =\frac{3\gann^2m_a^3}{8\pi}.
 \label{eq:nunuwidth}
\end{equation}
The photon operator gives
\begin{equation}
 \Gamma_{\gamma\gamma}
 =\frac{\gagg^2m_a^3}{64\pi}.
 \label{eq:ggwidth}
\end{equation}
With all other ALP operators switched off, the total width and invisible
branching fraction are
\begin{equation}
 \Gamma_a=\Gamma_{\nu\bar\nu}+\Gamma_{\gamma\gamma},
 \qquad
 \mathrm{BR}_{\rm inv}
 =\frac{\Gamma_{\nu\bar\nu}}{\Gamma_a}.
 \label{eq:brinv}
\end{equation}
The factors in Eq.~\eqref{eq:singleflavorwidth} correspond to the Dirac
particle--antiparticle convention used by the UFO. A Majorana realization may
require a different field normalization and identical-particle convention and
should not be inferred from these numbers without rebuilding the model.

\subsection{Mono-photon production}

\begin{figure}[tbp]
\centering
\includegraphics[width=.62\textwidth]{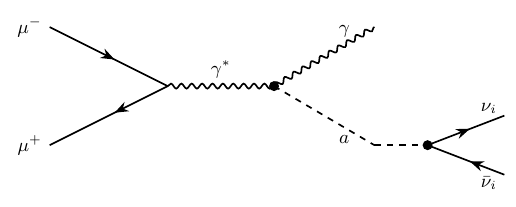}
\caption{Leading-order diagram for the mono-photon signal
$\mu^+\mu^-\to\gamma^*\to\gamma a$, followed by
$a\to\nu_i\bar\nu_i$. The production vertex is generated by the
$aF_{\mu\nu}\widetilde F^{\mu\nu}$ interaction.}
\label{fig:monofeynman}
\end{figure}

The generated signal is the flavor-summed channel
\begin{equation}
 \mu^+\mu^-\longrightarrow\gamma a,
 \qquad a\longrightarrow\nu_i\bar\nu_i,
 \quad i=e,\mu,\tau .
 \label{eq:signalprocess}
\end{equation}
The leading-order production topology is shown in figure~\ref{fig:monofeynman}.
The incoming muons annihilate through a virtual photon, which converts to the
observed photon and the ALP through the $a\gamma\gamma$ interaction; the ALP
then decays invisibly through its neutrino coupling.
At tree level, ALP production proceeds through an $s$-channel virtual photon,
$\mu^+\mu^-\to\gamma^*\to\gamma a$. Because the generated cards contain no
$aZ\gamma$ coupling (section~\ref{sec:ufo}), there is no $Z^*$-exchange
contribution and hence no $\gamma^*$--$Z^*$ interference. Neglecting the muon mass, the analytic
angular distribution in the normalization of Eq.~\eqref{eq:lagrangian} is
\begin{equation}
 \frac{d\sigma(\mu^+\mu^-\to\gamma a)}{d\cos\theta}
 =\frac{\alpha_{\rm em}\gagg^2}{64}
 \left(1-\frac{m_a^2}{s}\right)^3(1+\cos^2\theta),
 \label{eq:diffxsec}
\end{equation}
and hence
\begin{equation}
 \sigma(\mu^+\mu^-\to\gamma a)
 =\frac{\alpha_{\rm em}\gagg^2}{24}
 \left(1-\frac{m_a^2}{s}\right)^3.
 \label{eq:productionxsec}
\end{equation}
The cubic threshold factor is the characteristic two-body phase-space and
pseudoscalar-momentum suppression. In the narrow-width approximation,
\begin{equation}
 \sigma(\gamma+\not E)
 =\sigma(\mu^+\mu^-\to\gamma a)\,
 \mathrm{BR}(a\to\nu\bar\nu).
 \label{eq:nwaxsec}
\end{equation}
The largest benchmark width is $\Gamma_a/m_a\simeq1.07\%$, so factorization is
a useful approximation for the present inclusive rate. The matrix-element
results are about 2.7--2.8\% above Eq.~\eqref{eq:nwaxsec} across the full scan;
the stability of this offset is used as the normalization-closure test shown
later in Fig.~\ref{fig:closure}.

Since $\mathrm{BR}_{\rm inv}\simeq1$, the mono-photon rate is almost independent
of $y_\nu$ once the ALP decays promptly and invisibly. This explains the
near-degeneracy of BP2 and BP3: they have equal mass and photon coupling but
different neutrino couplings. Their widths differ, while their inclusive
mono-photon rates do not differ appreciably.

\subsection{Coupling rescaling and domain of validity}

The reference photon coupling is $\gagg=0.01\TeV^{-1}$. To leading order,
$\sigma(\mu^+\mu^-\to\gamma a)\propto\gagg^2$. More exactly, the invisible
rate obeys
\begin{equation}
 \sigma(\gamma+\not E)\propto
 \gagg^2\,
 \frac{\Gamma_{\nu\bar\nu}}
 {\Gamma_{\nu\bar\nu}+\Gamma_{\gamma\gamma}(\gagg)}.
 \label{eq:exactgscale}
\end{equation}
The contour construction neglects the mild coupling dependence of the second
factor and uses pure $\gagg^2$ scaling. This is numerically accurate here
because $\Gamma_{\nu\bar\nu}\gg\Gamma_{\gamma\gamma}$: even at the upper edges
of the plotted coupling ranges, the departure of $\mathrm{BR}_{\rm inv}$ from
unity is at most at the percent level. It would fail for much larger photon
couplings or smaller $y_\nu$.

The dimension-five photon operator also has a finite range of validity, as in
any collider ALP EFT analysis~\cite{Brivio:2017ije,Bauer:2020jbp}. A
naive scale inferred from the coefficient is $\Lambda_{\rm eff}\sim1/\gagg$,
up to UV-dependent coupling and normalization factors. Points for which
$\sqrts$ approaches or exceeds the actual mediator/compositeness scale cannot
be interpreted reliably within the contact-operator description. In
particular, the high-coupling portion of a 10 TeV contour should be accompanied
by an EFT-validity or perturbative-unitarity check in any precision claim.

\subsection{Generated samples and excluded topology}

Each of the ten benchmark--energy combinations contains 100,000 generated
signal events. Hard events are generated with
MadGraph5\_aMC@NLO~\cite{Alwall:2014hca}, showered with
Pythia~8~\cite{Sjostrand:2014zea}, and processed with the Delphes fast detector
simulation~\cite{deFavereau:2013fsa}. The
diphoton-plus-missing-energy topology is not a prediction of this minimal
massless-neutrino model: the radiation amplitude required for
$\mu^+\mu^-\to\nu\bar\nu a$, followed by $a\to\gamma\gamma$, is absent.
Raising $y_\nu$ or forcing the decay cannot supply the missing field content
or production vertex. We therefore introduce and analyze a distinct
phenomenological HNL realization in Appendix~\ref{app:hnlfeasibility}; its samples and
statistical interpretation are kept separate from the detector-level
mono-photon result.

The detector samples used for the numerical results were produced with the
generic Delphes card distributed through the MadGraph installation, with a
$3.8$ T solenoid, a $1.29$ m tracker radius and the calorimeter and object
parameterizations recorded in the run banners. This is not the dedicated
Muon Collider Detector ``TARGET'' card, whose response is itself explicitly a
target parameterization rather than a full-simulation validation
~\cite{MuonColliderDelphes}. Therefore detector-model dependence,
beam-induced backgrounds~\cite{Bartosik:2020} and shielding-cone losses are not included in the
quoted reach. In particular, the $|\eta_\ell|<2.5$ lepton veto below is a
central-veto assumption and does not reject forward $\mu^+\mu^-\gamma$ or
$e^+e^-\gamma$ events. A $10^\circ$ polar-angle boundary corresponds to
$|\eta|\simeq2.44$, so the central veto reaches only to the edge of the
shielding-nozzle region; no forward-lepton tagger is modeled here. The quoted
reach is consequently conditional on these omitted backgrounds being
controlled by forward instrumentation and is an optimistic baseline, not an
experiment-level projection. Beam-induced activity can be reduced with
timing selections in a dedicated detector simulation, but the present samples
contain neither beam overlays nor fake/merged-photon modeling; timing alone
does not determine their residual rate. An experiment-level projection
requires reprocessing with a versioned muon-collider card and beam-overlay
samples.

The five parameter points and their derived two-channel widths are summarized
in table~\ref{tab:benchmarks}. BP2 and BP3 deliberately share the same mass so
that the change in neutrino coupling can be separated from kinematic mass
effects.

\begin{table}[tbp]
\centering
\caption{Benchmark definitions and two-channel decay properties. The photon
coupling is $\gagg=0.01\TeV^{-1}$ for every point.}
\label{tab:benchmarks}
\small
\begin{tabular}{lrrrrrrr}
\toprule
BP & $m_a$ [GeV] & $y_\nu$ & $\gann$ [GeV$^{-1}$] & $\Gamma_{\nu\nu}$ [GeV] & $\Gamma_{\gamma\gamma}$ [GeV] & $\Gamma_a/m_a$ & BR$_\mathrm{inv}$\\
\midrule
BP1 & 100 & 0.10 & $1.000\times10^{-3}$ & 0.1194 & $4.974\times10^{-7}$ & 0.001194 & 0.999996\\
BP2 & 300 & 0.10 & $3.333\times10^{-4}$ & 0.3581 & $1.343\times10^{-5}$ & 0.001194 & 0.999963\\
BP3 & 300 & 0.30 & $1.000\times10^{-3}$ & 3.2229 & $1.343\times10^{-5}$ & 0.010743 & 0.999996\\
BP4 & 500 & 0.30 & $6.000\times10^{-4}$ & 5.3715 & $6.217\times10^{-5}$ & 0.010743 & 0.999988\\
BP5 & 1000 & 0.30 & $3.000\times10^{-4}$ & 10.7430 & $4.974\times10^{-4}$ & 0.010743 & 0.999954\\
\bottomrule
\end{tabular}
\end{table}

Table~\ref{tab:benchmarks} shows that BP2 and BP3 have the same ALP mass but different neutrino
couplings, allowing the effect of $y_\nu$ to be isolated. Because both points
already have an almost unit invisible branching fraction, their mono-photon
production rates remain very similar. BP1 and BP2 have $\Gamma_a/m_a\simeq
1.2\times10^{-3}$, while BP3--BP5 have $\Gamma_a/m_a\simeq1.07\times10^{-2}$;
this clustering follows directly from holding $y_\nu=m_a\gann$ fixed within
the BP1--BP2 and BP3--BP5 families, and all points remain in a relatively
narrow-width regime. In every benchmark the neutrino width exceeds the
diphoton width by several orders of magnitude, so the mono-photon rate probes
production through $\gagg$ followed by an essentially unit invisible
branching fraction.

\section{Event selection and statistical definitions}
\label{sec:selection}

The cut-based and TMVA workflows use related but not identical selections.
Their complete numerical definitions are collected in
table~\ref{tab:selections}; keeping them separate is essential when comparing
their quoted significances. For two-body production the photon energy is
\begin{equation}
 E_\gamma=\frac{s-m_a^2}{2\sqrts},\qquad
 x_\gamma\equiv\frac{2E_\gamma}{\sqrts}=1-\frac{m_a^2}{s},
\end{equation}
which motivates the mass-dependent line window in the cut-based analysis. The
cut-based requirement $|x_\gamma-(1-m_a^2/s)|<0.02$ is not applied to the TMVA
sample; instead, the recoil mass is supplied to the classifier. The background samples are
$\nu\bar\nu\gamma$ (irreducible), $WW\gamma$ with lost leptons, and a reducible
$\gamma\gamma$ control sample. Instrumental and beam-induced backgrounds are
not simulated. The veto is restricted to $|\eta_\ell|<2.5$ and thus leaves
forward $\mu^+\mu^-\gamma$ events, including their large $t$-channel
contribution, and forward $e^+e^-\gamma$ events untested. A forward muon
tagger (and corresponding forward electron/track veto) would be needed to
establish their rejection;
its efficiency cannot be inferred from the generic detector card used here.

The photon-line window also contains radiative return whenever
\begin{equation}
 |x_Z-x_a|=\frac{|m_a^2-m_Z^2|}{s}<0.02,\qquad
 x_Z=1-\frac{m_Z^2}{s}.
\end{equation}
At $3\TeV$, $x_Z=0.99908$ and the BP1 line is $x_a=0.99889$. In the
existing unweighted $\nu\bar\nu\gamma$ sample, after the photon
$p_T>0.05\sqrt{s}$ and $|\eta_\gamma|<2.5$ requirements, 72.6\% of the
BP1-window events have recoil mass within $5\GeV$ of $m_Z$ (65.9\% at
$10\TeV$). The line window therefore does not remove this resonant component;
the result includes its modeled contribution and is sensitive to the
recoil-mass and photon-energy response. This fraction is a parton-level
diagnostic before detector smearing, not a measured detector efficiency.

To expose the forward-lepton acceptance, we generated additional unweighted
parton-level $\mu^+\mu^-\gamma$ and $e^+e^-\gamma$ samples at $3\TeV$ and
$\mu^+\mu^-\gamma$ at $10\TeV$, with $|\eta_\ell|<10$ at generation (the
default MadGraph $|\eta_\ell|<2.5$ cut would remove the region under study).
Each sample contains 1000 events, with $p_T^\gamma>0.05\sqrt{s}$,
$|\eta_\gamma|<2.5$, and $\Delta R(\gamma,\ell)>0.4$. Table~\ref{tab:m4forward}
shows that extending the idealized lepton veto from $|\eta|<2.5$ to 3 reduces
the accepted $\mu^+\mu^-\gamma$ fraction by factors of 7.0 and 8.8 at 3 and
$10\TeV$, respectively. At $|\eta|<4$ no event survives in these pilot
samples. These are truth-level coverage scenarios, not validated tagging
efficiencies behind the shielding nozzle. No event survives the lepton veto in the BP1, BP4, or BP5 line window
in any generated pilot; with only 1000 events per sample, these zero counts
are limited-MC observations and do not establish a negligible residual.
The electron sample is available only at $3\TeV$; the $10\TeV$ electron
sample and higher-statistics samples remain necessary for a quantitative
background contribution to the sensitivity.

\begin{table}[tbp]
\centering
\caption{Forward-lepton veto scan in the additional parton-level pilot
samples. Fractions are before the ALP line window; every sample has 1000
unweighted events. The generated cross section includes the hard-photon
threshold and $|\eta_\gamma|<2.5$ acceptance.}
\label{tab:m4forward}
\small
\begin{tabular}{lrrr}
\toprule
Sample & $\sigma_{\rm gen}$ [fb] & survivors, $|\eta_\ell|<2.5$ & survivors, $|\eta_\ell|<3$ / $<4$ \\
\midrule
$\mu^+\mu^-\gamma$, $3\TeV$ & 142.1 & 105/1000 & 15/1000; 0/1000 \\
$\mu^+\mu^-\gamma$, $10\TeV$ & 15.77 & 88/1000 & 10/1000; 0/1000 \\
$e^+e^-\gamma$, $3\TeV$ & 1.941 & 0/1000 & 0/1000; 0/1000 \\
\bottomrule
\end{tabular}
\end{table}

Muon-collider studies identify beam-induced background as a major source of
out-of-time detector activity. In the tracker study of
Ref.~\cite{Bartosik:2020}, hit times are required to agree with the
interaction-point time of flight within $\pm3\sigma_T$, for assumed tracker
resolutions $\sigma_T=50$--$100\,\mathrm{ps}$; this mitigates out-of-time hit
occupancy. It does not determine the fake- or merged-photon rate in our
selection. Since this work has no beam-background overlay, we cannot assign
that residual yield or claim that a timing cut eliminates it. This
contribution and the forward-lepton backgrounds must be included before
interpreting the projected sensitivities as an experiment-level reach.

\begin{table}[tbp]
\centering
\caption{Complete numerical event selections. The cut-based workflow and the
TMVA preselection are intentionally reported in separate columns because they
are not identical. The balance variable is
$A_{p_T}=|p_T^\gamma-\met|/(p_T^\gamma+\met)$.}
\label{tab:selections}
\scriptsize
\begin{tabular}{p{0.18\textwidth}p{0.36\textwidth}p{0.36\textwidth}}
\toprule
Requirement & Cut-based analysis & TMVA preselection\\
\midrule
Photon candidate & leading photon & leading photon\\
Candidate $p_T$ & $p_T^\gamma>0.05\sqrts$ & $p_T^\gamma\geq0.02\sqrts$\\
Missing momentum & $\met>0.05\sqrts$ & $\met\geq0.02\sqrts$\\
Photon acceptance & $|\eta_\gamma|<2.5$ & $|\eta_\gamma|\leq2.5$\\
Lepton veto & no $e/\mu$ with $p_T>10$ GeV, $|\eta|<2.5$ & no $e/\mu$ with $p_T\geq10$ GeV, $|\eta|\leq2.5$\\
Extra photons & exactly one photon with $p_T>0.02\sqrts$, $|\eta|<2.5$ & no explicit extra-photon veto\\
Azimuthal recoil & $\Delta\phi(\gamma,\met)>2.8$ & $\Delta\phi(\gamma,\met)\geq2.4$\\
$p_T$ balance & not applied & $A_{p_T}\leq0.50$\\
Mass information & $|x_\gamma-(1-m_a^2/s)|<0.02$ & recoil mass used as MVA input\\
\bottomrule
\end{tabular}
\end{table}

At 3 (10) TeV the cut-based 5\% photon and missing-momentum threshold is 150
(500) GeV, while the TMVA 2\% threshold is 60 (200) GeV. The accepted-photon
threshold used only for the cut-based extra-photon veto is likewise 60 (200)
GeV. Table~\ref{tab:selections} therefore provides all numerical object,
kinematic, angular, balance, veto, and line-window requirements used in the
reported results.

For the cut-based study the approximate significance is
\begin{equation}
 Z_\mathrm{cut}=\frac{S}{\sqrt{S+B+(\delta_BB)^2}},\qquad \delta_B=0.01.
 \label{eq:zcut}
\end{equation}
For the held-out TMVA test sample, the implementation uses
\begin{equation}
 Z_\mathrm{MVA}=\frac{S}{\sqrt{B+(\delta_BB)^2}}.
 \label{eq:zmva}
\end{equation}
The numerical difference is negligible when $S\ll B$, as is true here, but the
two definitions are recorded explicitly for reproducibility. If a simulated
background component has zero passing events, it is not treated as exactly
zero: a three-event, approximately 95\% Poisson upper-count floor is propagated
through the event weight. This prescription is a conservative regularization,
not a substitute for a finite-Monte-Carlo nuisance parameter. For the primary
BDTG result the optimized irreducible-background tails contain only 2--41
unweighted test events, corresponding to naive relative counting errors
$1/\sqrt{n}=16$--71\%. Results based on fewer than 25 passing events are
therefore explicitly regarded as tail-statistics limited. The exact passing
counts are retained in the machine-readable
\texttt{background\_post\_mva.csv} files.

At a reference coupling $g_0=0.01\TeV^{-1}$, the projected signal yield is
\begin{equation}
 S(g)=S(g_0)\left(\frac{g}{g_0}\right)^2.
 \label{eq:gscale}
\end{equation}
The quoted $g_{95}$ and $g_{5\sigma}$ values solve $Z=1.64$ and $Z=5$,
respectively. They are sensitivity thresholds, not observed limits. A rigorous
profile-likelihood treatment with nuisance parameters would follow the methods
of refs.~\cite{Cowan:2010js,Read:2002hq}; it is not performed here.

\section{Cut-based analysis results}

\begin{figure}[tbp]
\centering
\includegraphics[width=.61\textwidth]{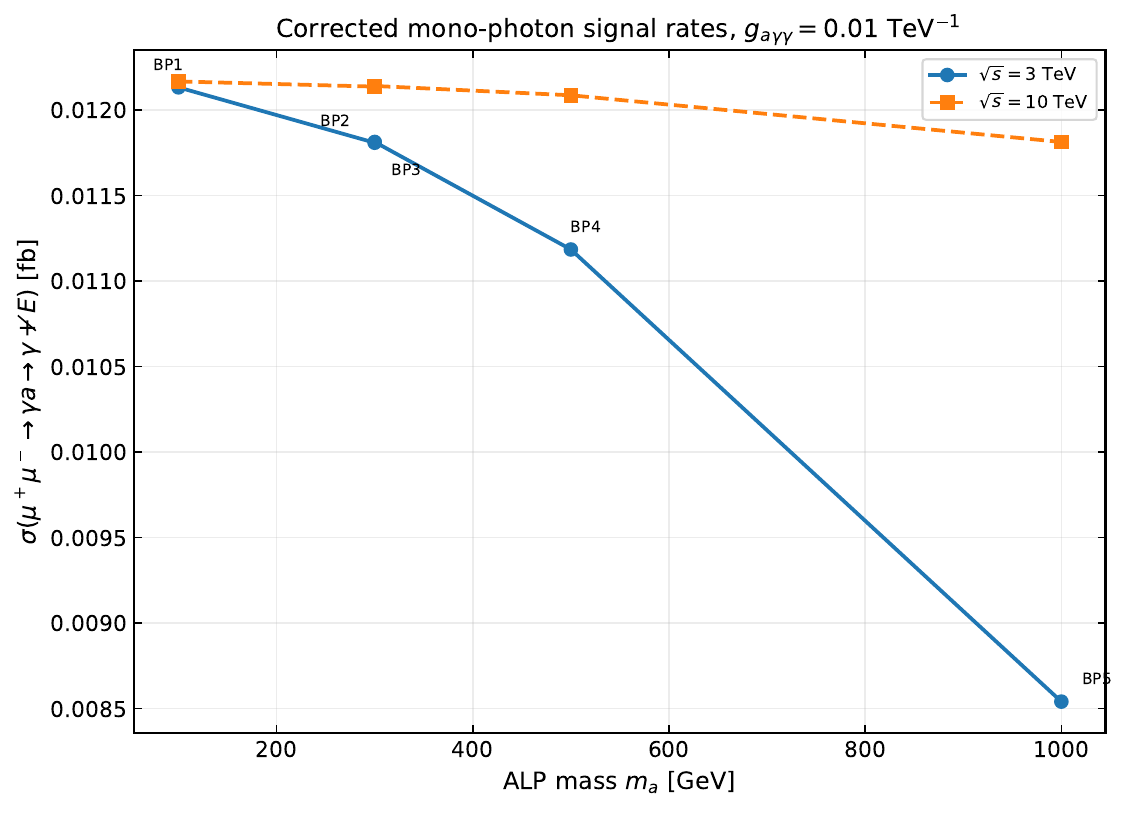}
\caption{Mono-photon signal cross section versus ALP mass at the
reference photon coupling.}
\label{fig:xsec}
\end{figure}

Figure~\ref{fig:xsec} shows that the signal rates are of order
$10^{-2}$ fb. The 3 and 10 TeV rates are nearly equal for light and
intermediate masses, reflecting the high-energy behavior of the production
amplitude in this range, so the tenfold luminosity at 10 TeV yields roughly
ten times more raw signal events. At 3 TeV the BP5 rate decreases because the
$m_a=1$ TeV state occupies a noticeably smaller fraction of the available
phase space. The figure shows generated benchmark nodes joined for visual
guidance; the lines are not an independent continuous mass scan.

\begin{figure}[tbp]
\centering
\includegraphics[width=.88\textwidth]{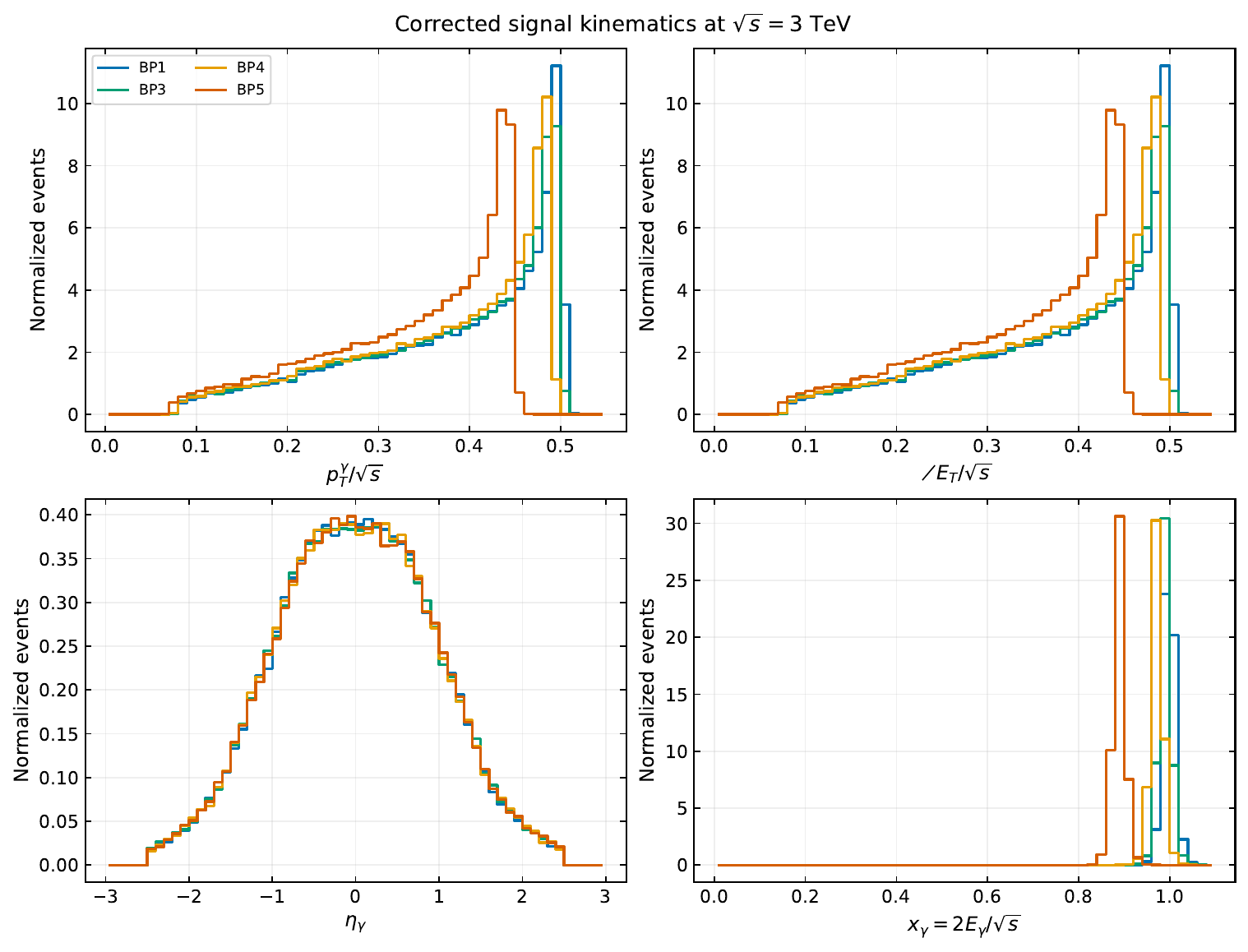}
\caption{Normalized signal kinematics at $\sqrts=3\TeV$.}
\label{fig:kin3}
\end{figure}

In figure~\ref{fig:kin3}, the photon-$p_T$ and missing-$p_T$ spectra closely track one
another because the leading-order final state recoils as a two-body system.
The $x_\gamma$ distribution forms a mass-dependent line near
$1-m_a^2/s$; the displacement is largest for BP5. The pseudorapidity panel
describes detector acceptance and is much less directly tied to the ALP mass.
These correlations explain why a line window is powerful and why redundant
variables should be avoided in the multivariate classifier.

\begin{figure}[tbp]
\centering
\includegraphics[width=.88\textwidth]{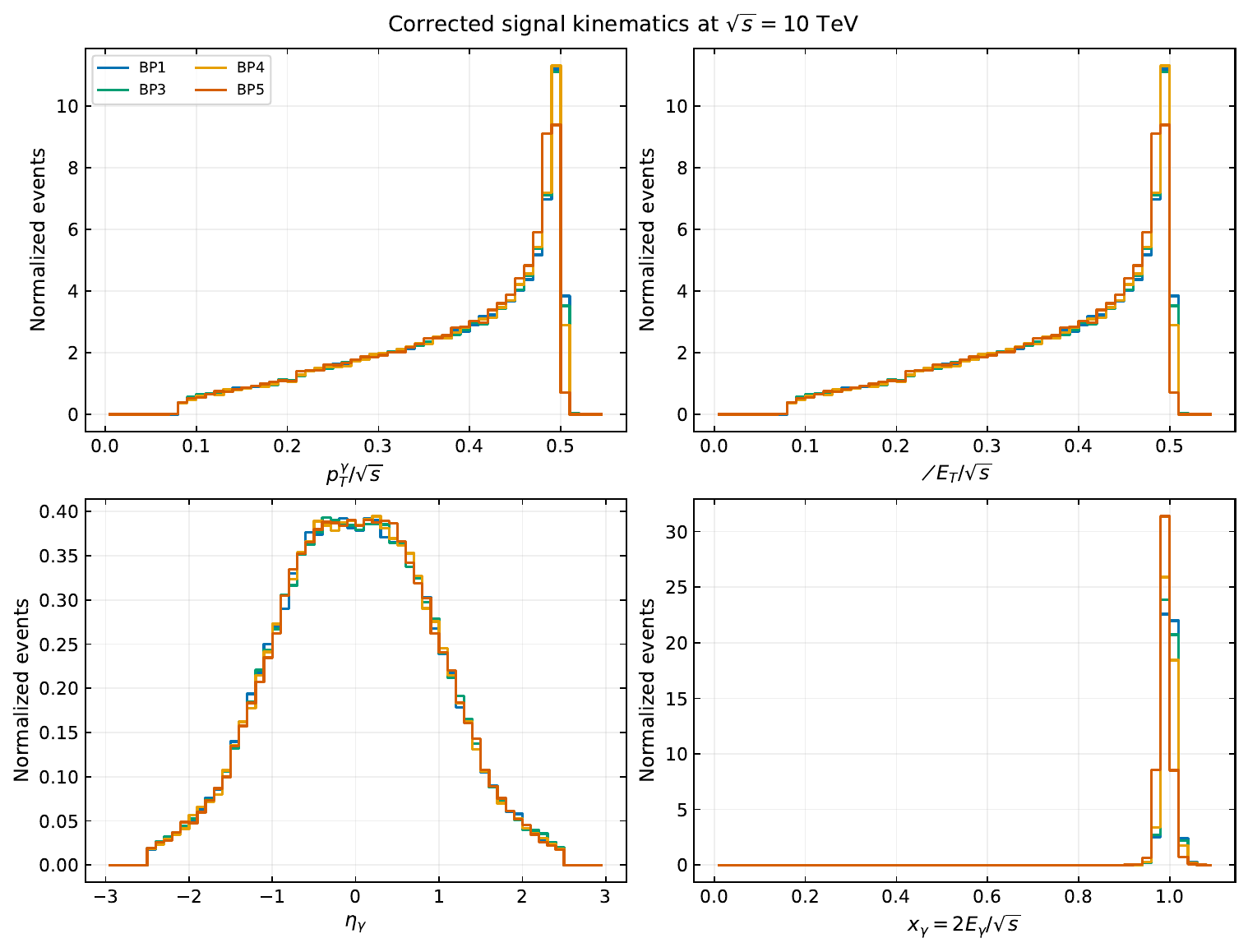}
\caption{Normalized signal kinematics at $\sqrts=10\TeV$.}
\label{fig:kin10}
\end{figure}

Figure~\ref{fig:kin10} shows that at 10 TeV, even a 1 TeV ALP has $m_a^2/s=0.01$, so all
benchmarks cluster closer to the high-$x_\gamma$ endpoint than at 3 TeV. This
compresses the mass-dependent separation in the photon line while preserving
the characteristic back-to-back topology. The larger luminosity, rather than
a dramatically larger production cross section, drives the improved event
yield at 10 TeV.

\begin{figure}[tbp]
\centering
\includegraphics[width=.88\textwidth]{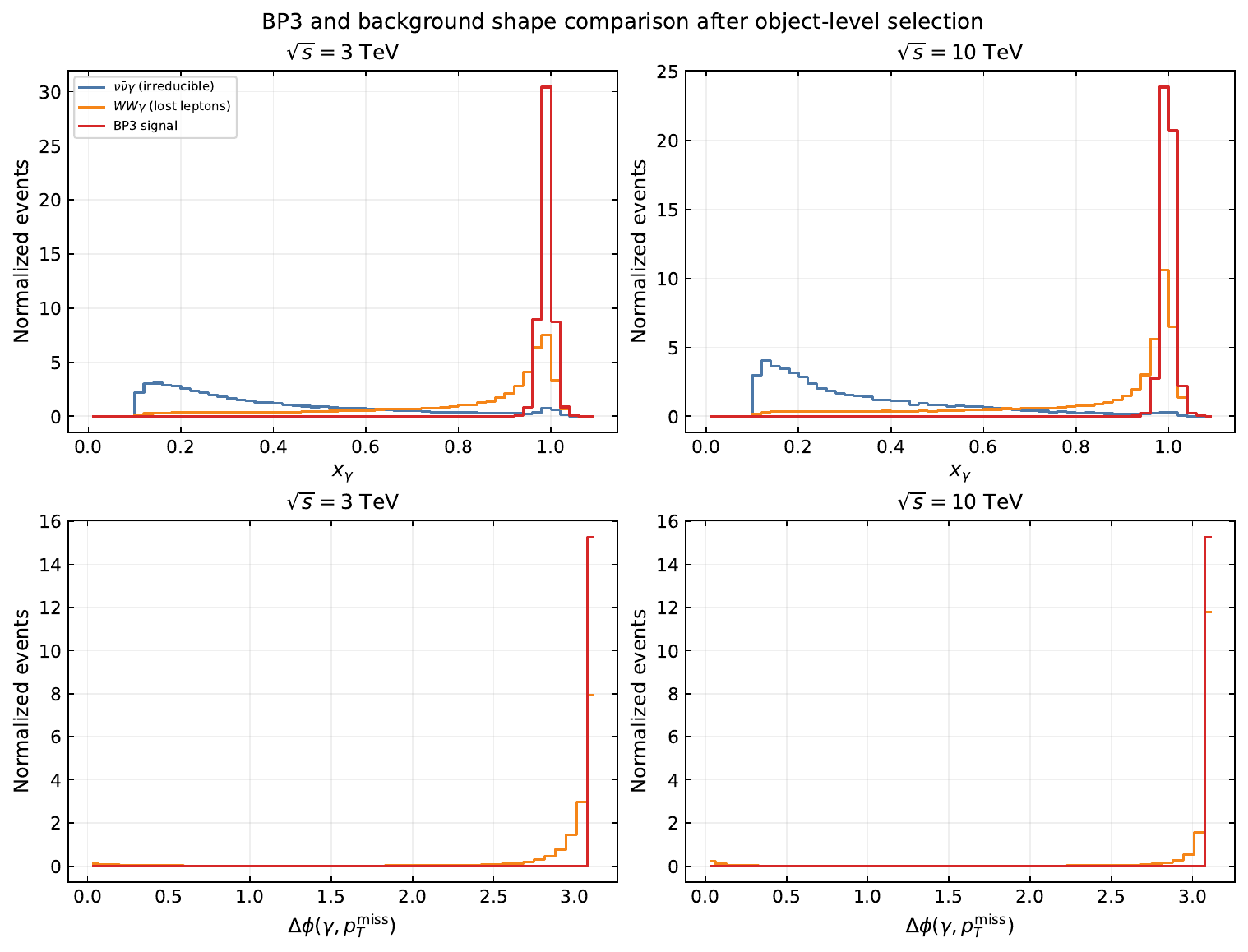}
\caption{BP3 signal and background shapes after object-level selection.}
\label{fig:shapes}
\end{figure}
Figure~\ref{fig:shapes} shows that both the signal and irreducible background favor a
back-to-back configuration, so $\Delta\phi(\gamma,\met)$ is primarily a broad
topology-cleaning variable. The sharper discrimination comes from
$x_\gamma$: the signal is localized by two-body kinematics, whereas the
$\nu\bar\nu\gamma$ and lost-lepton backgrounds are continuous. The reducible
diphoton control has no post-veto shape because it is fully removed in the
available sample.
\begin{figure}[tbp]
\centering
\includegraphics[width=.88\textwidth]{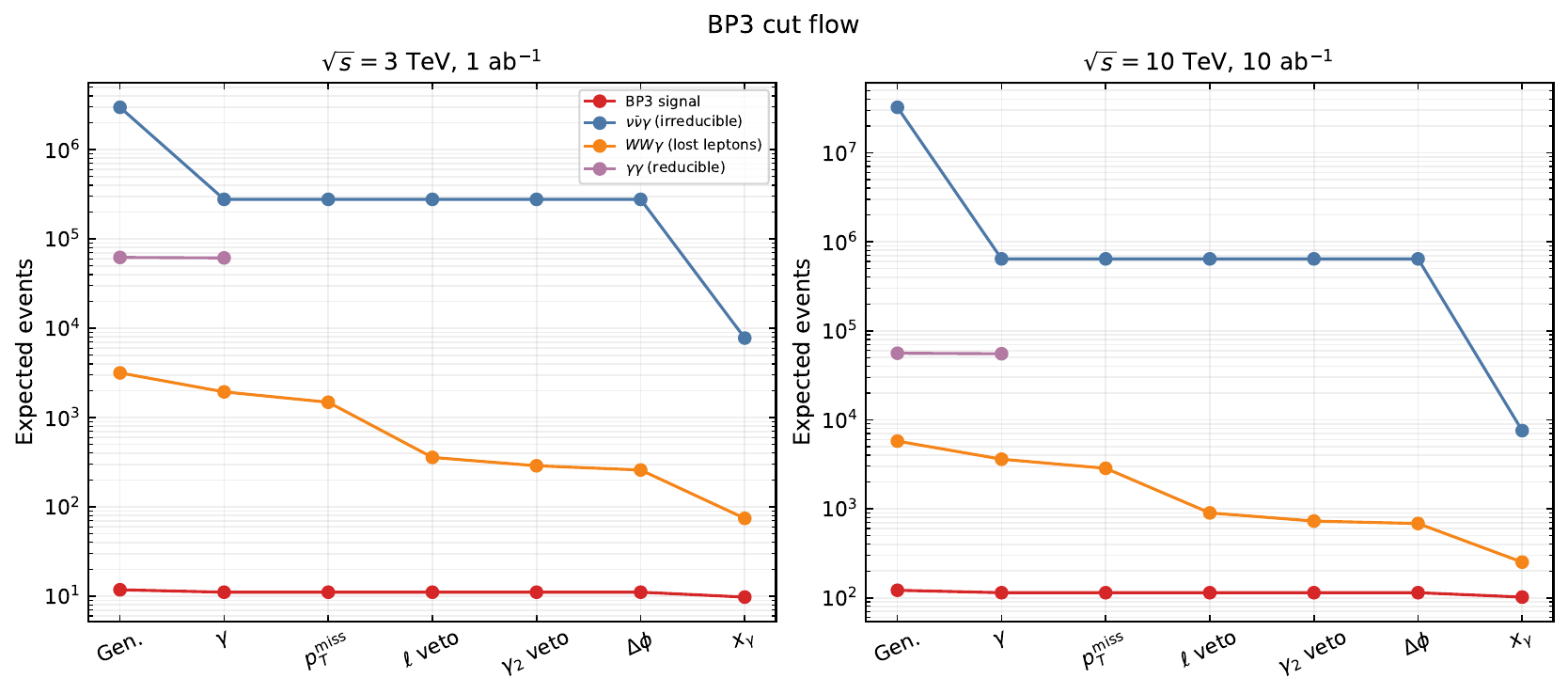}
\caption{Weighted BP3 cut flow at the two collider energies.}
\label{fig:cutflow}
\end{figure}
Figure~\ref{fig:cutflow} shows that the initial background is many orders of magnitude larger
than the signal. Photon acceptance and missing momentum remove the reducible
diphoton sample, while the lepton veto suppresses $WW\gamma$. The final
photon-line requirement produces the largest reduction of the irreducible
$\nu\bar\nu\gamma$ component while retaining about 83\% of signal. Even after
this reduction, the analysis remains background dominated.
\begin{table}[tbp]
\centering
\caption{Final cut-based yields, background composition and approximate
sensitivity. The background $B$ is split into its $\nu\bar\nu\gamma$,
$WW\gamma$ and $\gamma\gamma$ components; the $\gamma\gamma$ entry is the
weighted three-event upper-count floor.}
\label{tab:cutsensitivity}
\footnotesize
\setlength{\tabcolsep}{3pt}
\begin{tabular}{llrrrrrrrrrr}
\toprule
 & & & & \multicolumn{4}{c}{$B$} & & & & \\
\cmidrule(lr){5-8}
BP&$\sqrts$&$\epsilon_S$&$S$&$\nu\bar\nu\gamma$&$WW\gamma$&$\gamma\gamma$ floor&Total&$S/B$&$Z_\mathrm{stat}$&$\Zsys$&$g_{95}$\\
 & [TeV]&&&&&&&&&&[TeV$^{-1}$]\\
\midrule
BP1&3&0.8273&10.04&7841&63.0&0.186&7904&0.00127&0.113&0.0843&0.04424\\
BP1&10&0.8359&101.69&7391&248.5&0.168&7640&0.01331&1.156&0.8727&0.01373\\
BP2&3&0.8289&9.79&7763&74.5&0.186&7838&0.00125&0.110&0.0827&0.04467\\
BP2&10&0.8357&101.42&7553&251.3&0.168&7805&0.01300&1.141&0.8572&0.01385\\
BP3&3&0.8284&9.78&7763&74.5&0.186&7838&0.00125&0.110&0.0827&0.04467\\
BP3&10&0.8385&101.79&7553&251.3&0.168&7805&0.01304&1.145&0.8603&0.01383\\
BP4&3&0.8321&9.31&5016&72.5&0.186&5088&0.00183&0.130&0.1061&0.03948\\
BP4&10&0.8375&101.21&7715&255.7&0.168&7971&0.01270&1.127&0.8427&0.01397\\
BP5&3&0.8312&7.10&3182&17.9&0.186&3200&0.00222&0.125&0.1091&0.03899\\
BP5&10&0.8400&99.22&7553&263.6&0.168&7817&0.01269&1.115&0.8378&0.01402\\
\bottomrule
\end{tabular}
\end{table}

Table~\ref{tab:cutsensitivity} shows that, at the generated coupling, none of the cut-based points reaches
the nominal 95\% target. The 10 TeV values are substantially larger because of
the tenfold luminosity, but the 1\% background systematic reduces the gain
relative to a statistics-only estimate. The inferred cut-based reach is
$g_{95}\simeq0.039$--0.045 TeV$^{-1}$ at 3 TeV and approximately
$0.014\TeV^{-1}$ at 10 TeV. The irreducible $\nu\bar\nu\gamma$ process
controls the final uncertainty, while the lost-lepton component is subleading
but non-negligible. The diphoton floor has no numerical impact on the totals,
yet retaining it prevents a finite Monte Carlo sample with zero survivors from
being interpreted as proof of zero background. The BP dependence of $B$ arises
from the mass-dependent photon-line window, and this background admitted by
each window controls the modest variation of the sensitivity among masses
more than the small variation in signal cross section.

\begin{longtable}{llrrrrr}
\caption{BP3 cut flow in expected event yields.}\label{tab:bp3flow}\\
\toprule $\sqrts$&Selection&Signal&$\nu\bar\nu\gamma$&$WW\gamma$&$\gamma\gamma$&Total $B$\\\midrule
\endfirsthead
\toprule $\sqrts$&Selection&Signal&$\nu\bar\nu\gamma$&$WW\gamma$&$\gamma\gamma$&Total $B$\\\midrule
\endhead
3&Generated&11.81&2,976,738.9&3165.3&62,136.0&3,042,040.2\\
3&Photon acceptance&11.10&277,554.1&1940.2&61,329.8&340,824.2\\
3&Missing-energy cut&11.10&277,554.1&1485.5&0&279,039.6\\
3&Lepton veto&11.10&277,554.1&358.3&0&277,912.4\\
3&Extra-photon veto&11.10&277,554.1&288.8&0&277,842.9\\
3&Back-to-back&11.10&277,554.1&259.0&0&277,813.1\\
3&Photon-line window&9.78&7763.3&74.5&0&7837.8\\\midrule
10&Generated&121.39&32,416,673.5&5754.6&55,924.0&32,478,352.1\\
10&Photon acceptance&114.01&642,466.1&3605.8&55,202.7&701,274.6\\
10&Missing-energy cut&114.01&642,466.1&2849.3&0&645,315.3\\
10&Lepton veto&114.01&642,466.1&898.5&0&643,364.5\\
10&Extra-photon veto&114.01&642,466.1&727.6&0&643,193.6\\
10&Back-to-back&114.01&642,466.1&683.1&0&643,149.1\\
10&Photon-line window&101.79&7553.1&251.3&0&7804.4\\
\bottomrule
\end{longtable}

Table~\ref{tab:bp3flow} gives the numerical cut flow and confirms the visual conclusion from
Fig.~\ref{fig:cutflow}. The photon-line selection suppresses the remaining
irreducible background by factors of about 36 at 3 TeV and 85 at 10 TeV while
retaining approximately 88\%--89\% of the signal that survives acceptance.
The table lists the nominal zero diphoton yield; the three-event floor is added
only when the final sensitivity is computed.

\begin{figure}[tbp]
\centering
\includegraphics[width=.71\textwidth]{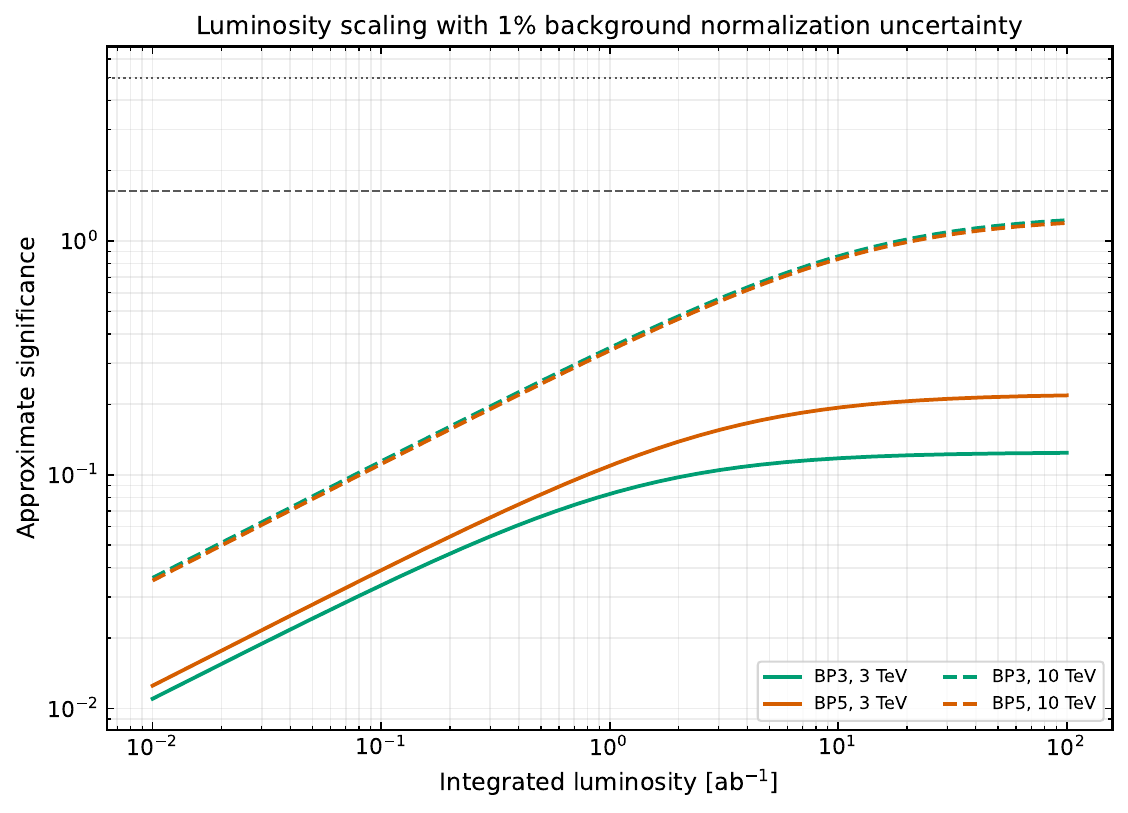}
\caption{Luminosity scaling for BP3 and BP5 with a 1\% background uncertainty.}
\label{fig:lumi}
\end{figure}
In figure~\ref{fig:lumi}, at low luminosity the curves grow approximately as
$\sqrt{\mathcal L}$. Once $(0.01B)^2$ dominates the denominator, the gain
saturates toward $S/(0.01B)$. Therefore luminosity alone cannot indefinitely
compensate for weak $S/B$; improved background rejection or reduced
normalization uncertainty becomes essential.
\begin{figure}[tbp]
\centering
\includegraphics[width=.98\textwidth]{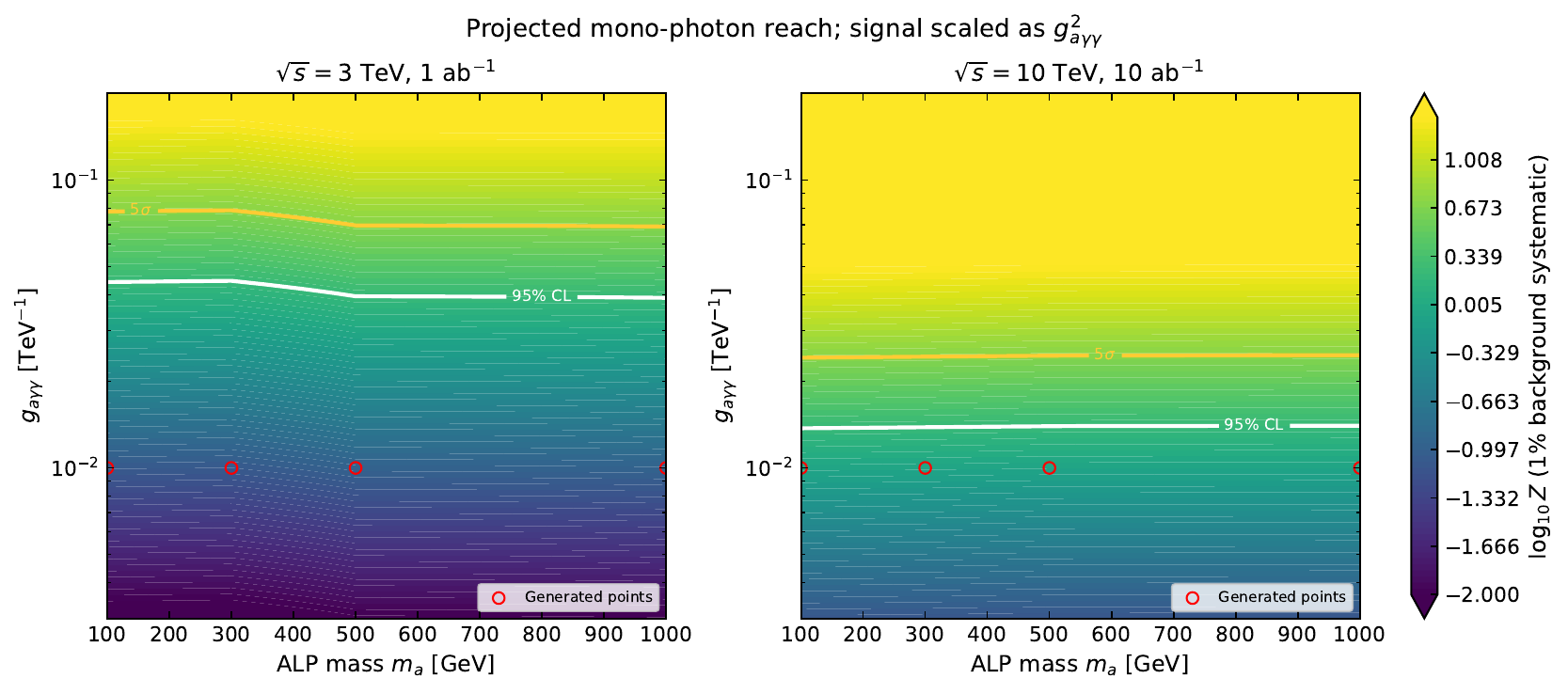}
\caption{Cut-based projected mass--photon-coupling sensitivity.}
\label{fig:cutcontour}
\end{figure}
In figure~\ref{fig:cutcontour}, the white and yellow curves denote the approximate $Z=1.64$
and $Z=5$ thresholds. Only the red mass nodes were simulated. Between those
nodes, signal and background yields are linearly interpolated in mass; away
from the generated coupling, only the signal normalization is rescaled with
Eq.~\eqref{eq:gscale}. Acceptance, widths, branching ratios, and the cut
optimization are held fixed. Hence the colored plane is a projected
sensitivity surface in the sense of section~\ref{sec:selection}, not a direct
two-dimensional simulation.
\begin{figure}[tbp]
\centering
\includegraphics[width=0.71\textwidth]{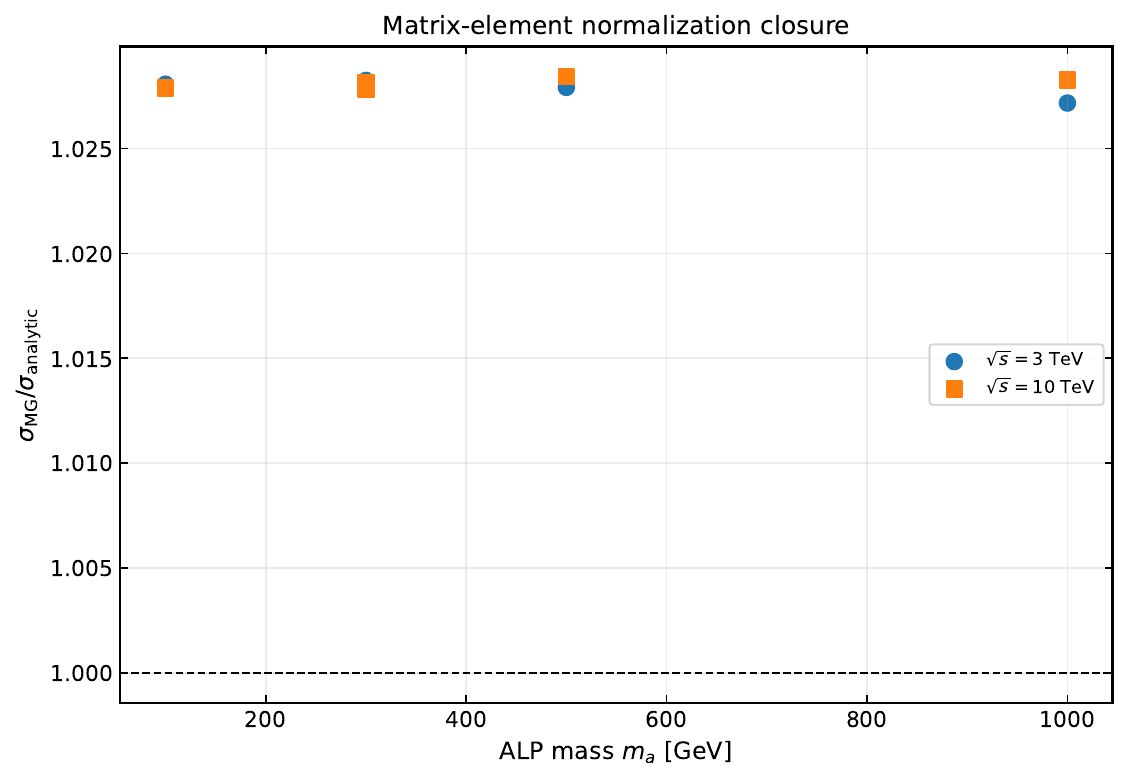}
\caption{MadGraph-to-analytic cross-section closure.}
\label{fig:closure}
\end{figure}
Figure~\ref{fig:closure} shows that the ratio remains close to 1.028 for every benchmark and both
energies. Its stability supports the internal mass and coupling dependence of
the generated samples. The residual few-percent offset may reflect the input
electromagnetic scheme and Monte Carlo integration conventions; it should be
treated as a normalization check rather than tuned away point by point.

\section{TMVA analysis and held-out classifier results}

The multivariate analysis is implemented with ROOT/TMVA~\cite{Brun:1997pa,
Hocker:2007ht}. BDTG is designated as the primary classifier throughout this
paper because its regularized tree ensemble is robust for the non-Gaussian
recoil tails. MLP and Likelihood results are diagnostic method-variation
checks and are not used to form a best-of-methods envelope. All classifiers use identical deterministic
60\%/20\%/20\% training, validation, and held-out test partitions. The MLP has
one hidden layer with $N+5$ nodes, and the Likelihood method uses spline
probability-density estimates. The input observables are leading-photon
$p_T$, $|\eta_\gamma|$, and recoil mass. Missing transverse momentum and
transverse mass are omitted because they are nearly deterministic copies of
photon $p_T$ in the exclusive two-body signal. Class weights balance signal
and background for training while retaining the physical relative mixture of
background processes. The decision threshold is selected only on validation
data, then applied once to the held-out test sample. This prescription avoids
selecting a different classifier for each benchmark
after examining the held-out sensitivity. The complete held-out results for
all three classifiers are listed in table~\ref{tab:allmva} of
appendix~\ref{app:mvaresults}.

The finite size of the held-out tails is quantified in
table~\ref{tab:bdtgtails}. The quoted $1/\sqrt{n}$ column is a diagnostic of
the unweighted irreducible-background count only; it is not added in
quadrature to the 1\% normalization uncertainty in the central values. Nine
of the ten operating points have fewer than 25 passing events. Consequently,
the BDTG reach is a central projection with an explicitly identified
Monte-Carlo-tail limitation, not a precision limit.

\begin{table}[tbp]
\centering
\caption{Finite-sample diagnostic for the primary BDTG operating points. Here
$n_{\nu\nu\gamma}$ is the unweighted number of held-out irreducible-background
events passing the validation-fixed threshold, and $B$ is the total weighted
background used in the significance.}
\label{tab:bdtgtails}
\small
\begin{tabular}{lrrrrr}
\toprule
BP & $\sqrts$ [TeV] & $n_{\nu\nu\gamma}$ & $1/\sqrt{n}$ [\%] & $B$ & Tail limited?\\
\midrule
BP1&3&41&15.6&3054&no\\ BP2&3&12&28.9&906.1&yes\\
BP3&3&6&40.8&454.5&yes\\ BP4&3&3&57.7&238.2&yes\\
BP5&3&2&70.7&156.5&yes\\ BP1&10&7&37.8&6007&yes\\
BP2&10&7&37.8&6070&yes\\ BP3&10&6&40.8&5171&yes\\
BP4&10&7&37.8&6005&yes\\ BP5&10&3&57.7&2612&yes\\
\bottomrule
\end{tabular}
\end{table}

Table~\ref{tab:bdtgtails} shows why a high AUC does not by
itself establish a precise reach. The large event weights convert a handful of
surviving $\nu\bar\nu\gamma$ events into hundreds or thousands of expected
events. Additional background simulation is therefore more urgent than
further classifier tuning.

\begin{figure}[tbp]
\centering
\includegraphics[width=.88\textwidth]{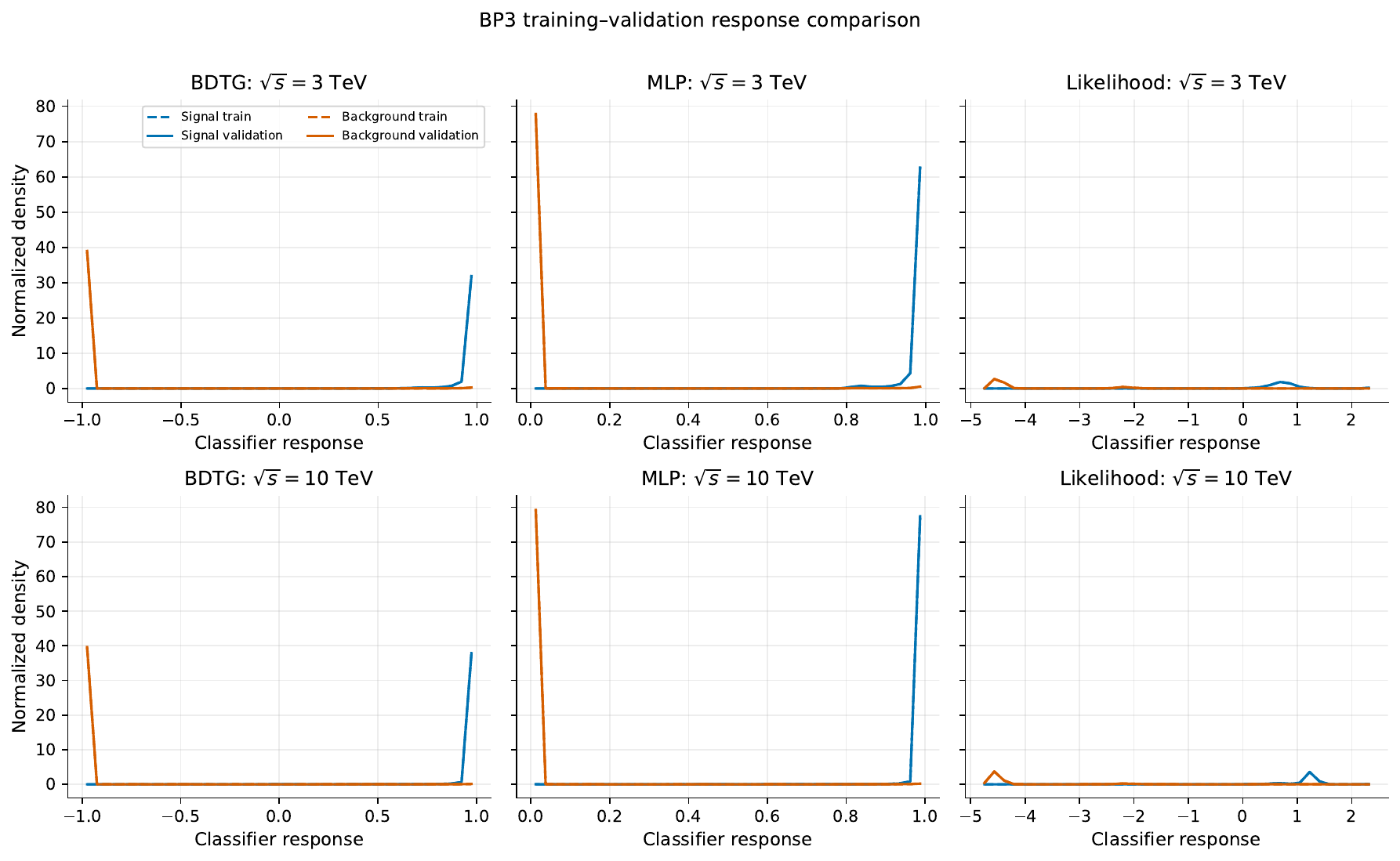}
\caption{BP3 classifier-response distributions for the training and validation
samples.}
\label{fig:overtrain}
\end{figure}

In figure~\ref{fig:overtrain}, for each method and energy, the solid validation distributions
track the dashed training distributions closely. The maximum
training--validation cumulative histogram distances are all below 0.008,
providing no visible evidence of severe overtraining. Likelihood has the
largest discrepancies, up to 0.0079 for the 10 TeV background, consistent with
its more variable tail behavior. This is a diagnostic, not a formal
goodness-of-fit probability: the metric compares binned cumulative shapes,
does not incorporate the look-elsewhere effect or systematic variations, and
should be supplemented by stability studies if the classifier configuration
is changed.

\begin{figure}[tbp]
\centering
\includegraphics[width=.88\textwidth]{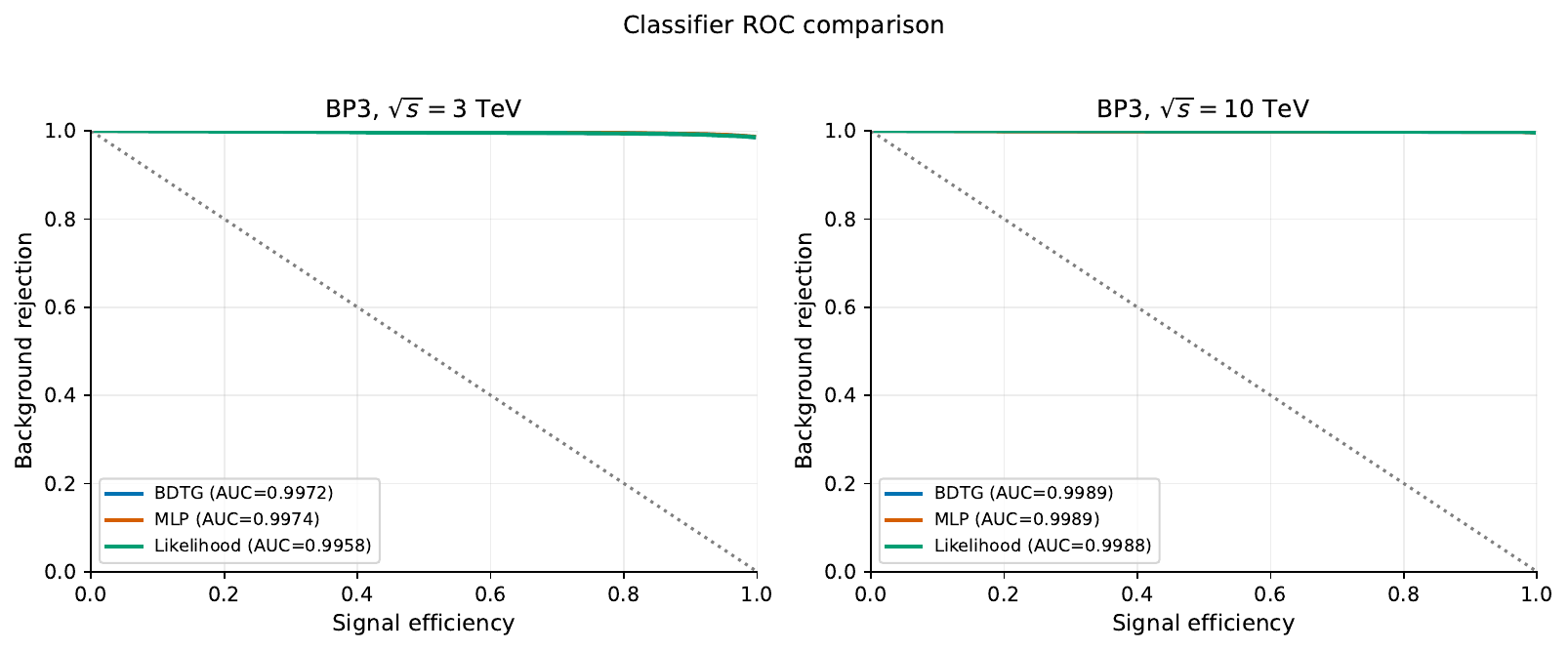}
\caption{Validation-sample ROC comparison for BP3.}
\label{fig:roc}
\end{figure}

Figure~\ref{fig:roc} shows that all three methods achieve validation AUC values above 0.995,
with BDTG and MLP slightly ahead of Likelihood at 3 TeV. At 10 TeV the AUC
values are almost indistinguishable. AUC measures ranking over all thresholds,
however, whereas the physics significance depends on one threshold and on
event weights. Similar AUC values therefore do not guarantee similar final
sensitivity.

\begin{figure}[tbp]
\centering
\includegraphics[width=.88\textwidth]{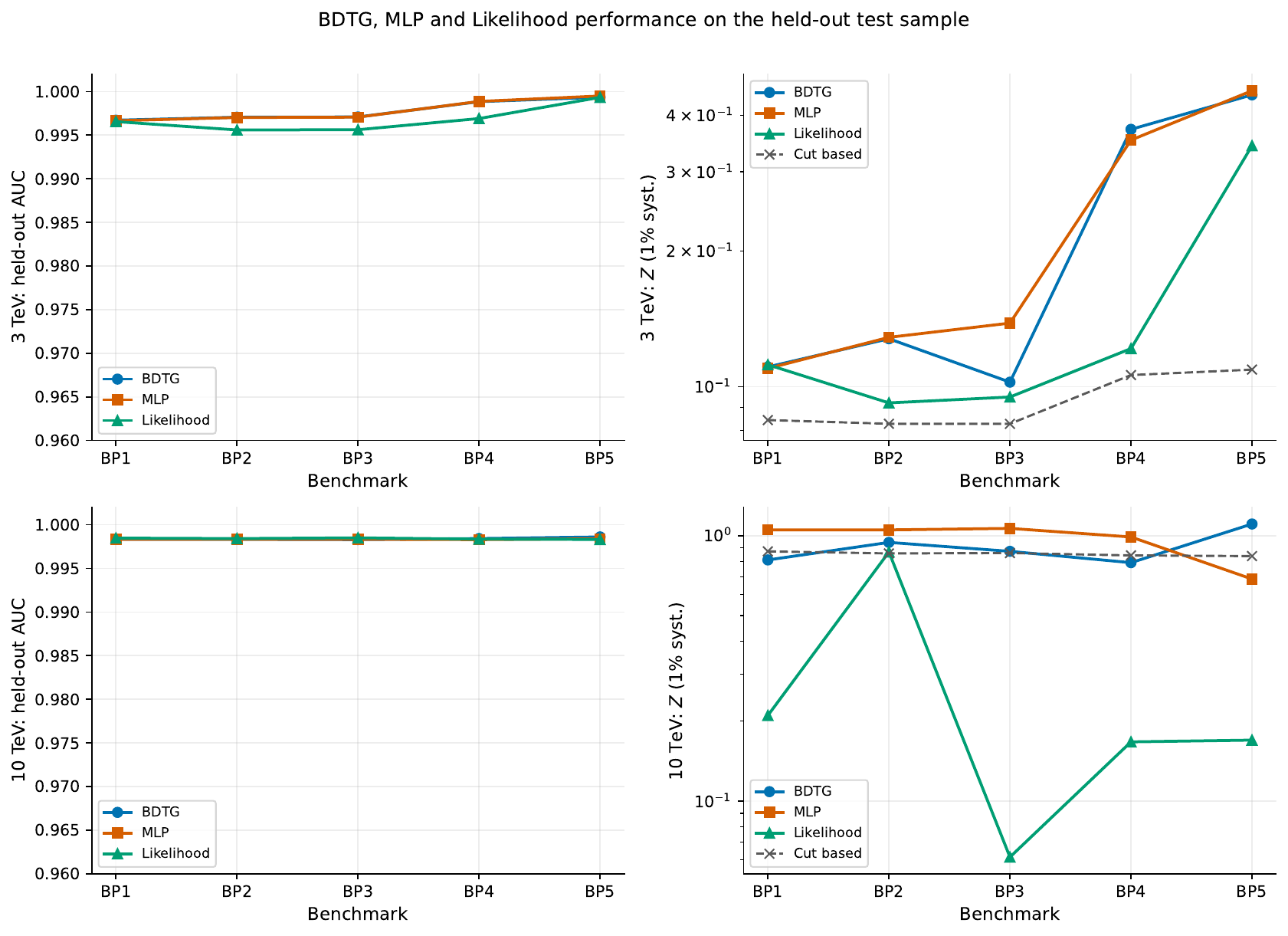}
\caption{Held-out AUC and significance for the three classifiers, compared with
the cut-based result.}
\label{fig:performance}
\end{figure}

Figure~\ref{fig:performance} shows that the test AUC remains close to validation performance across all
benchmarks. The MVA improves significance most clearly for heavier points at
3 TeV and for several 10 TeV benchmarks. The MLP variation happens to exceed
BDTG for some points, but this post-test ordering is not used to select the
reported primary method. The Likelihood result is unstable in physics
sensitivity despite high AUC---especially BP3 at 10 TeV---because its
validation-selected threshold retains too little signal and is sensitive to
the weighted background tail.

\begin{figure}[tbp]
\centering
\includegraphics[width=.87\textwidth]{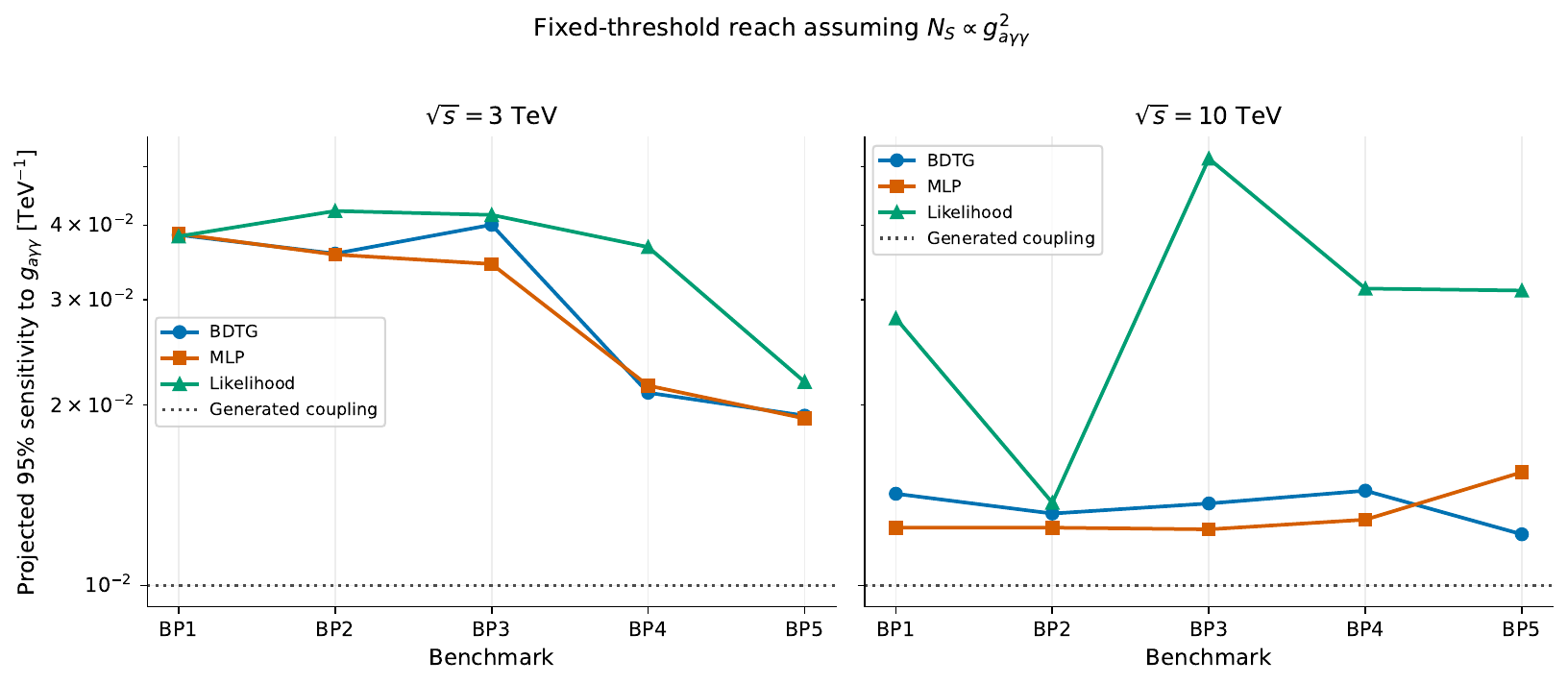}
\caption{Approximate fixed-threshold coupling reach for each classifier.}
\label{fig:mvareach}
\end{figure}

Figure~\ref{fig:mvareach} shows that the primary BDTG gives
$g_{95}=0.0192$--$0.0400\TeV^{-1}$ at 3 TeV and
$0.0122$--$0.0144\TeV^{-1}$ at 10 TeV. The MLP curve is a method-variation
diagnostic rather than a source of point-by-point replacements. The much weaker
Likelihood values for several 10 TeV points reflect the operating threshold,
not poor global ranking. The generated reference coupling lies below the
projected 95\% threshold for every method and point.

\begin{figure}[tbp]
\centering
\includegraphics[width=\textwidth]{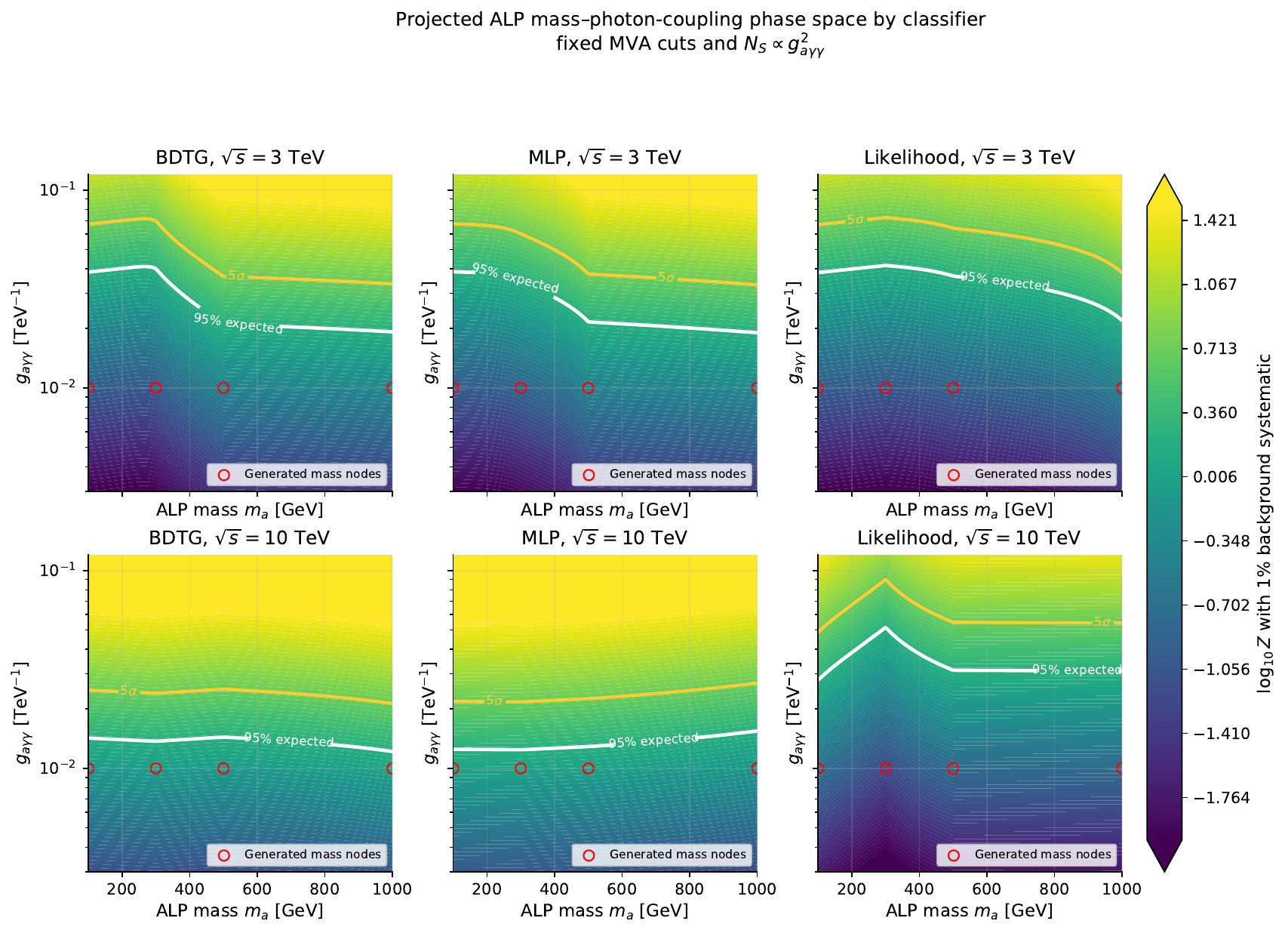}
\caption{Classifier-dependent projected ALP mass--photon-coupling phase space.}
\label{fig:mvacontour}
\end{figure}

\begin{figure}[tbp]
\centering
\includegraphics[width=.82\textwidth]{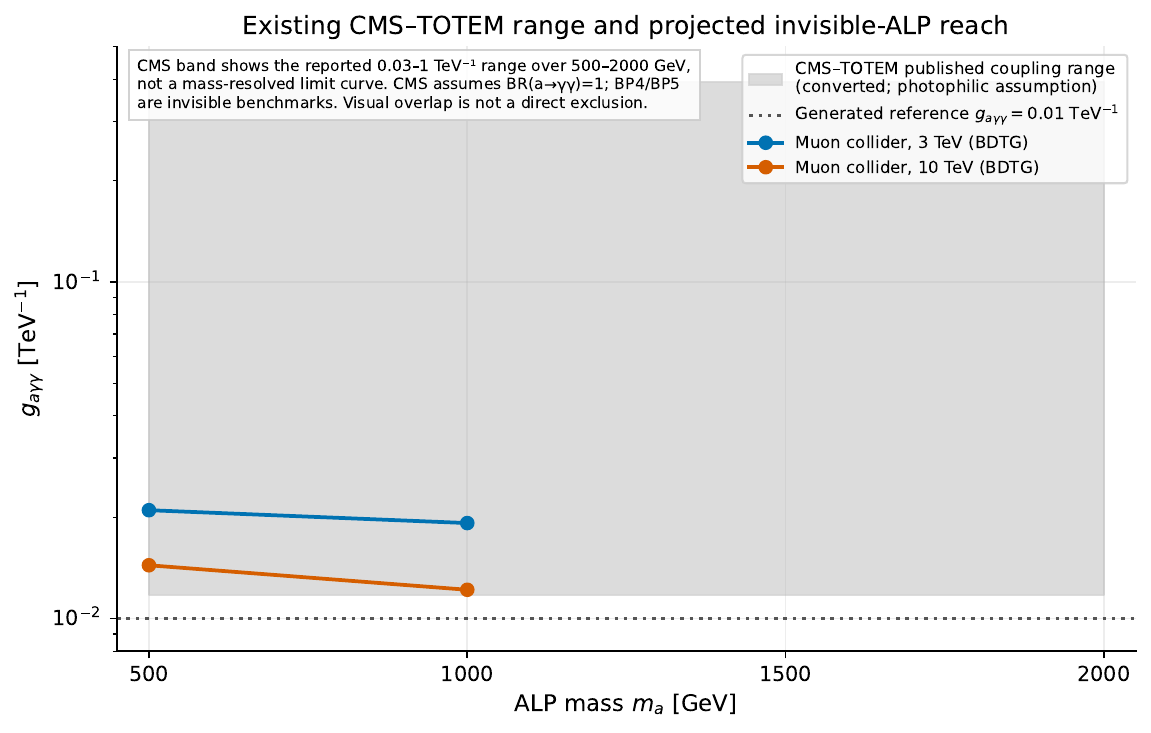}
\caption{Converted CMS--TOTEM photophilic coupling range and projected BDTG
sensitivities for the invisible-ALP benchmarks. The gray region is the full
range $f^{-1}=0.03$--$1\TeV^{-1}$ reported over the mass interval, converted
using $g_{a\gamma\gamma}=4e^2f^{-1}$; it is not a mass-resolved limit curve.
CMS assumes $\mathrm{BR}(a\to\gamma\gamma)=1$, unlike BP4/BP5.}
\label{fig:external_limits}
\end{figure}

Each panel of figure~\ref{fig:mvacontour} uses only BP1, BP3, BP4, and BP5 as independent mass
nodes; BP2 is omitted because it shares $m_a=300$ GeV with BP3. Signal and
background yields are interpolated linearly in mass, and signal is scaled as
$\gagg^2$ with the MVA cut fixed. The corresponding BP1, BP3, BP4 and BP5
entries of table~\ref{tab:allmva} are therefore the only independently
analyzed mass coordinates entering the figure; BP2 is a valuable coupling
benchmark but cannot be a second independent mass node at 300 GeV. The
corresponding $5\sigma$ values remain in the machine-readable result tables
and follow $g_{5\sigma}=g_{95}\sqrt{5/1.64}$ only under the fixed-background
scaling used here. The differences between panels quantify
method dependence; they are not a substitute for retraining and
regenerating samples throughout the two-dimensional parameter plane.

\paragraph{Separate HNL feasibility study.} A benchmark-level, parton-level HNL diphoton exercise is documented in Appendix~\ref{app:hnlfeasibility}; it is outside the mono-photon model and does not contribute to the sensitivities reported here.

\section{Comparison with related ALP--neutrino scenarios}
\label{sec:modelcomparison}

Table~\ref{tab:modelcomparison} separates three physically different uses of
the phrase ``ALP--neutrino model.'' The BP3 number of this work is a generated
mono-photon rate in the low-energy pseudoscalar model of
Eq.~\eqref{eq:lagrangian}. The Grid-2 row refers to an archived sample, labelled
``Grid-2'' in our run records, that was generated with an earlier
implementation of the model; its rate is recovered directly from the archived
run banners. That sample assumes massless neutrinos and reports
$\nu\bar\nu a$ with $a\to\gamma\gamma$. That implementation did not contain a
consistent nonzero massless-neutrino radiation amplitude, so its tiny rate is
retained only for provenance and must not be reweighted into a constraint.
It is shown because it documents that a massless-neutrino implementation
cannot supply the diphoton topology, which motivates the HNL extension of
Appendix~\ref{app:hnlfeasibility}.
Finally, a genuine heavy-neutral-lepton (HNL) theory has extra parameters and
different production and decay chains. The published LHC example in
ref.~\cite{Alves:2019xpc} demonstrates this model class but does not provide a
cross section at the present muon-collider benchmark. Appendix~\ref{app:hnlfeasibility} records a dedicated 16-point, parton-level feasibility scan based on the HeavyN electroweak sector and phenomenological ALP vertices.

\begin{landscape}
\begin{table}[p]
\centering
\caption{Comparison of the present mono-photon result, an archived
massless-neutrino sample, and heavy-neutrino interpretations. Cross sections are quoted only when a
corresponding generated banner exists. Dashes therefore carry physical and
provenance information; they do not denote zero.}
\label{tab:modelcomparison}
\scriptsize
\begin{tabular}{@{}L{0.16\textwidth}L{0.20\textwidth}L{0.28\textwidth}rrL{0.24\textwidth}@{}}
\toprule
Scenario & Channel & Reference point & $\sigma(3\TeV)$ [fb] &
$\sigma(10\TeV)$ [fb] & Status and valid comparison\\
\midrule
This work, BP3 & $\gamma a$, $a\to\nu\bar\nu$ &
$m_a=300\GeV$, $y_\nu=0.30$, $\gagg=0.01\TeV^{-1}$ &
$1.181\times10^{-2}$ & $1.214\times10^{-2}$ &
Validated generated rate; three massless final-state neutrino flavors in the
declared phenomenological pseudoscalar model.\\
Archived Grid-2 sample & $\nu\bar\nu a$, $a\to\gamma\gamma$ &
$m_a=300\GeV$, $\gann=10^{-3}\GeV^{-1}$,
$\gagg=1\TeV^{-1}$ & $3.01\times10^{-11}$ &
$7.77\times10^{-12}$ & Massless-neutrino sample from an earlier UFO;
retained for provenance only and excluded from all sensitivities.\\
This work, explicit HNL scan & $\nu_\mu\bar\nu_\mu a$, $a\to\gamma\gamma$ &
$m_a=50$--$300\GeV$, $m_N=200$--$1000\GeV$, $|V_{\mu N}|=10^{-2}$,
$c_N=1$, $f_N=1\TeV$ & $0.1268$--$0.3392$ & $0.1308$--$0.3423$ &
Generated 16-point scan; parton-level feasibility results in
Appendix~\ref{app:hnlfeasibility}, separate from the detector-level mono-photon
projection.\\
Published sterile-neutrino ALP model~\cite{Alves:2019xpc} & LHC mono-photon
from a heavy-scalar cascade & 13 TeV $pp$ collider; light ALP, sterile-neutrino
and heavy-scalar parameters & --- & --- & Establishes a consistent UV/EFT model
class, but not a matched muon-collider benchmark.\\
Invisible-ALP muon-collider study~\cite{Casarsa:2021rud} & $\gamma a$ with
escaping $a$ & 3 (10) TeV and 1 (10) ab$^{-1}$ &
\multicolumn{2}{c}{$\Lambda_{95}=112\ (375)\TeV$} & Closest published
topology; its portal normalization and invisible-particle assumptions differ,
so $\Lambda_{95}$ is not identified with $1/\gagg$ here.\\
\bottomrule
\end{tabular}
\end{table}
\end{landscape}

The table prevents two misleading comparisons. First, the BP3 and
Grid-2 rows use different final states and photon couplings, so their numerical
ratio is not a physics enhancement factor. Second, neither a light-neutrino
coupling nor a single HNL mass specifies an HNL prediction: mixing, chirality,
widths and decay branching fractions are indispensable inputs. The HNL scan
reported here therefore remains a benchmark-specific feasibility study and
is not a direct sensitivity comparison with either the mono-photon analysis
or the broader ALP--HNL study of ref.~\cite{BurgosMarcos:2025}.

\section{Conclusions and outlook}
\label{sec:conclusions}

We have quantified the sensitivity of multi-TeV muon colliders to the
invisible ALP process $\mu^+\mu^-\to\gamma a\to\gamma\nu\bar\nu$ in a
minimal two-coupling model. For the five benchmarks with
$m_a=100$--$1000\GeV$, $y_\nu=0.1$ or $0.3$, and $\gagg=0.01\TeV^{-1}$, the
ALP decays invisibly with $\mathrm{BR}(a\to\nu\bar\nu)>0.99995$, and the
MadGraph mono-photon cross sections of $0.00854$--$0.01217\,\mathrm{fb}$
agree with the analytic two-body prediction up to a stable 2.7--2.8\%
normalization offset.

The cut-based photon-line selection retains $82.7$--$84.0\%$ of the signal,
but the selected sample remains dominated by the irreducible
$\nu\bar\nu\gamma$ background. Assuming a 1\% uncertainty on the background
normalization, it gives $g_{95}=0.039$--$0.045\TeV^{-1}$ at $3\TeV$ and
$0.0137$--$0.0140\TeV^{-1}$ at $10\TeV$. The predeclared primary BDTG
confirms that the line-like photon and recoil-mass kinematics contain strong
discriminating information: with held-out AUC values of $0.9967$--$0.9993$,
it reaches $g_{95}=0.0192$--$0.0400\TeV^{-1}$ at $3\TeV$ and
$0.0122$--$0.0144\TeV^{-1}$ at $10\TeV$. The MLP and Likelihood results
quantify algorithm dependence but are not mined for a point-by-point best
result.

The CMS--TOTEM exclusive-diphoton result also probes $m_a=500$--$2000\GeV$,
overlapping BP4 and BP5. Its quoted coupling is $f^{-1}=C_{\gamma\gamma}/\Lambda$;
in the normalization used here the post-EWSB photon operator is
$e^2 f^{-1}aF_{\mu\nu}\widetilde F^{\mu\nu}$, whereas
Eq.~\eqref{eq:lagrangian} uses $\gagg aF_{\mu\nu}\widetilde
F^{\mu\nu}/4$. Thus the coefficient translation is
$\gagg=4e^2f^{-1}=16\pi\alpha f^{-1}$ (approximately
$0.39f^{-1}$ using the running electromagnetic coupling at these scales).
The reference $\gagg=0.01\TeV^{-1}$ corresponds to
$f^{-1}\simeq0.0255\TeV^{-1}$ in the CMS convention, below the smallest
coupling value (about $0.03\TeV^{-1}$) reached by that search. At the two
shared mass nodes, the BDTG projections are $g_{95}=0.0210$ and
$0.0144\TeV^{-1}$ for BP4 at 500 GeV, and $0.0192$ and
$0.0122\TeV^{-1}$ for BP5 at 1000 GeV (3 and 10 TeV, respectively).
These values locate our projected reach relative to the published coupling
span, but the span alone is insufficient to derive a mass-by-mass improvement.
This conversion puts the published photophilic CMS--TOTEM limit on the same
operator-coefficient axis, but does not by itself make the exclusions directly
applicable to our benchmarks: CMS assumes $\mathrm{BR}(a\to\gamma\gamma)=1$,
whereas BP4 and BP5 have $\mathrm{BR}(a\to\nu\bar\nu)=0.999988$ and
$0.999954$, respectively. Consequently, the CMS diphoton limit does not
exclude these invisible-decay benchmarks without a reinterpretation including
their suppressed diphoton branching fraction. Any overlay of the two results
must label the CMS curve as the photophilic limit and the muon-collider points
as invisible-channel projected sensitivities; a ratio of their coupling
values is only a comparison of nominal reach, not an improvement in a common
signal model.

The $10\TeV$, $10\,\mathrm{ab}^{-1}$ stage therefore offers the stronger
projected coupling reach, but none of the generated
$\gagg=0.01\TeV^{-1}$ benchmarks reaches the one-sided $Z=1.64$ criterion;
the largest value is $Z=1.107$. The optimized BDTG irreducible-background
tails contain only 2--41 unweighted events, so their 16--71\% counting
uncertainty is an important limitation beyond the assumed 1\% normalization
systematic. The mass--coupling curves must consequently be read as projected
expected sensitivities, not observed exclusions. They use four independent
mass nodes, fixed classifier thresholds, linear interpolation of $S$ and $B$,
and $S\propto\gagg^2$ rescaling. They do not include coupling-dependent
retraining, width and acceptance changes, beam-induced or fake-photon
backgrounds, detector and theory shape systematics, or correlated nuisance
parameters.

A separate parton-level HNL feasibility exercise is documented in Appendix~\ref{app:hnlfeasibility}; it does not support a detector-level sensitivity claim. The archived massless-neutrino diphoton sample remains invalid as a substitute for that HNL topology. Within these stated assumptions, the study
identifies mono-photon recoil as a promising phenomenological probe, while
the sparse background tails and generic detector card preclude an
experiment-level exclusion claim.

\section*{Data and code availability}

The supplementary archive accompanying this article,
\path{JHEP_REPRODUCIBILITY_RELEASE.tar.gz}, contains the manuscript source,
publication figures, derived numerical tables, plot inputs, benchmark
manifests, analysis scripts, and the M4 forward-lepton pilot scan and its
three generated parton-level samples. The larger event-level MadGraph samples and
ROOT/TMVA files are available from the corresponding author on reasonable
request.

\section*{Acknowledgments}

During preparation of this manuscript, the authors used AI assistants (OpenAI Codex and
Anthropic Claude) for language editing, consistency checks and assistance with analysis-code and
\LaTeX{} revisions. All scientific choices, numerical validation and the final
text were reviewed by the authors, who take full responsibility for the
content.

\appendix
\section{Complete held-out classifier results}
\label{app:mvaresults}

Table~\ref{tab:allmva} lists the held-out results for all three classifiers
and all benchmark--energy combinations shown in figures~\ref{fig:performance}
and~\ref{fig:mvareach}. The large AUC values establish strong global
separation, but $Z$ is limited by the absolute weighted background in the
selected tail. Coupling reach follows $g_{95}=g_0\sqrt{1.64/Z(g_0)}$ for the
TMVA definition in Eq.~\eqref{eq:zmva}.

\footnotesize
\begin{longtable}{lllrrrrrr}
\caption{All held-out classifier results. $S$ and $B$ are expected yields at
the design luminosity; $B$ includes zero-count floors.}\label{tab:allmva}\\
\toprule BP&$\sqrts$&Method&AUC$_\mathrm{val}$&AUC$_\mathrm{test}$&$S$&$B$&$Z$&$g_{95}$\\\midrule
\endfirsthead
\toprule BP&$\sqrts$&Method&AUC$_\mathrm{val}$&AUC$_\mathrm{test}$&$S$&$B$&$Z$&$g_{95}$\\\midrule
\endhead
BP1&3&BDTG&0.9964&0.9967&6.99&3054&0.111&0.03850\\
BP1&3&Likelihood&0.9963&0.9966&8.82&4333&0.112&0.03828\\
BP1&3&MLP&0.9966&0.9967&7.85&3726&0.110&0.03865\\
BP2&3&BDTG&0.9972&0.9971&4.02&906.1&0.128&0.03582\\
BP2&3&Likelihood&0.9955&0.9956&8.69&5679&0.0921&0.04221\\
BP2&3&MLP&0.9974&0.9970&3.04&530.8&0.129&0.03570\\
BP3&3&BDTG&0.9972&0.9971&2.23&454.5&0.102&0.04003\\
BP3&3&Likelihood&0.9958&0.9956&8.14&4933&0.0949&0.04158\\
BP3&3&MLP&0.9974&0.9971&4.36&908.6&0.138&0.03442\\
BP4&3&BDTG&0.9988&0.9988&5.82&238.2&0.373&0.02098\\
BP4&3&Likelihood&0.9969&0.9969&5.87&1954&0.122&0.03674\\
BP4&3&MLP&0.9989&0.9989&5.49&236.8&0.353&0.02156\\
BP5&3&BDTG&0.9989&0.9993&5.60&156.5&0.444&0.01921\\
BP5&3&Likelihood&0.9989&0.9993&5.31&234.2&0.343&0.02187\\
BP5&3&MLP&0.9993&0.9995&5.71&155.9&0.454&0.01901\\
BP1&10&BDTG&0.9989&0.9983&79.5&6007&0.810&0.01423\\
BP1&10&Likelihood&0.9989&0.9985&6.35&841.7&0.210&0.02793\\
BP1&10&MLP&0.9989&0.9983&82.8&4323&1.05&0.01249\\
BP2&10&BDTG&0.9989&0.9983&93.2&6070&0.943&0.01318\\
BP2&10&Likelihood&0.9989&0.9984&67.8&4269&0.868&0.01375\\
BP2&10&MLP&0.9989&0.9983&82.8&4328&1.05&0.01249\\
BP3&10&BDTG&0.9989&0.9983&77.3&5171&0.873&0.01371\\
BP3&10&Likelihood&0.9988&0.9985&3.44&2502&0.0615&0.05163\\
BP3&10&MLP&0.9989&0.9983&83.9&4331&1.07&0.01241\\
BP4&10&BDTG&0.9990&0.9984&77.6&6005&0.791&0.01440\\
BP4&10&Likelihood&0.9988&0.9984&9.37&2511&0.167&0.03132\\
BP4&10&MLP&0.9989&0.9983&87.8&5193&0.989&0.01288\\
BP5&10&BDTG&0.9991&0.9986&63.5&2612&1.11&0.01217\\
BP5&10&Likelihood&0.9991&0.9983&9.51&2511&0.170&0.03108\\
BP5&10&MLP&0.9991&0.9984&53.4&4250&0.687&0.01545\\
\bottomrule
\end{longtable}

\section{Parton-level HNL diphoton feasibility study}
\label{app:hnlfeasibility}

\subsection{Mass-basis HNL realization and generated samples}

To obtain a nonzero diphoton-plus-missing-momentum signal without assigning an
unsuppressed pseudoscalar coupling to a massless chiral neutrino, we extend the
official \href{https://cp3.irmp.ucl.ac.be/projects/feynrules/wiki/HeavyN}
{Dirac HeavyN implementation}~\cite{Degrande:2016aje,Pascoli:2018heg} by one ALP. The electroweak $WN\ell$, $ZN\nu$,
Higgs and Goldstone vertices
are inherited unchanged from that implementation. Only the first Dirac HNL,
$N\equiv N_1$, mixes with the muon flavor; $V_{eN}=V_{\tau N}=0$ and all
mixings of $N_{2,3}$ are set to zero. The added phenomenological mass-basis
interactions are
\begin{align}
 {\cal L}_{aN}={}&-i y_N a\bar N\gamma_5N
 -i y_{N\nu}a\left(\bar\nu_\mu\gamma_5N+\bar N\gamma_5\nu_\mu\right)
 +\frac{\gagg}{4}aF_{\mu\nu}\widetilde F^{\mu\nu},
 \label{eq:hnllagrangian}\\
 y_N=\frac{c_Nm_N}{f_N},\qquad &y_{N\nu}=y_NV_{\mu N}.
 \label{eq:hnlcouplings}
\end{align}
This is a useful collider realization but is not by itself a gauge-invariant
ultraviolet completion of the ALP sector. In particular, the added vertices
were inserted directly in the UFO rather than derived from one common
FeynRules Lagrangian. The HNL total width includes the standard
$N\to\mu W,\nu_\mu Z,\nu_\mu h$ modes and, when open,
\begin{equation}
 \Gamma(N\to\nu_\mu a)=\frac{|y_{N\nu}|^2m_N}{16\pi}
 \left(1-\frac{m_a^2}{m_N^2}\right)^2,
 \qquad
 \Gamma(a\to\gamma\gamma)=\frac{\gagg^2m_a^3}{64\pi}.
 \label{eq:hnlwidths}
\end{equation}

\begin{figure}[tbp]
\centering
\includegraphics[width=.98\textwidth]{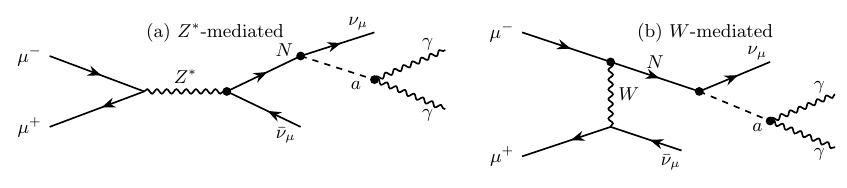}
\caption{Representative leading-order contributions to
$\mu^+\mu^-\to\nu_\mu\bar\nu_\mu a\to
\nu_\mu\bar\nu_\mu\gamma\gamma$: (a) neutral-current production through
$Z^*$ and (b) charged-current production through $W$ exchange. In both
topologies the intermediate HNL $N_1$ radiates the ALP, which decays through the
$a\gamma\gamma$ vertex. Charge-conjugate diagrams are omitted from the
drawing.}
\label{fig:hnlfeynman}
\end{figure}

We generate the inclusive process in Eq.~\eqref{eq:hnlprocess} with MadGraph5\_aMC@NLO using\footnote{The process card reads \texttt{generate mu+ mu- > vm vm\textasciitilde{} ax / n2 n2\textasciitilde{} n3 n3\textasciitilde{} QED=2 NP=1, ax > a a NP=1}. The exclusion applies only to $N_{2,3}$; the finite-width $N_1$ propagator, including its on-shell pole, is retained.} the full matrix element with on-shell $N_1$ allowed:
\begin{equation}
 \mu^+\mu^-\to\nu_\mu\bar\nu_\mu a,
 \qquad a\to\gamma\gamma,
 \label{eq:hnlprocess}
\end{equation}
Representative $Z$- and $W$-mediated contributions are shown in
figure~\ref{fig:hnlfeynman}. In each, an intermediate $N_1$ emits the ALP through
the added $a\nu_\mu N_1$ interaction, and the ALP decays to two photons;
the charge-conjugate contributions are also included in the generated
matrix element. The figure illustrates representative topologies rather than
the complete set of diagrams.
The grid uses $m_a=50,100,300\GeV$ and $m_N=200,500,1000\GeV$, retaining
only $m_N>m_a$. All 16 mass--energy points contain 50,000 unweighted events
and use $|V_{\mu N}|=10^{-2}$, $c_N=1$, $f_N=1\TeV$ and
$\gagg=1\TeV^{-1}$. The reference $g_{a\gamma\gamma}=1\TeV^{-1}$ is a simulation benchmark. LEP $Z$-pole searches constrain the photon-only operator below $m_Z$~\cite{Jaeckel:2015jla}, and ATLAS/CMS light-by-light searches cover ALP masses up to about 100 GeV~\cite{ATLAS:2020lbyl,CMS:2024lbyl}. For the operator normalization $\mathcal L\supset(g_{a\gamma\gamma}/4)aF\widetilde F$ used here, the inverse-scale convention $1/\Lambda_a$ is $g_{a\gamma\gamma}$. These light-ALP limits do not exclude the 50 and 100 GeV points at the reference coupling. The 300 GeV points are covered by the ATLAS forward-proton diphoton search over 150--1600 GeV~\cite{ATLAS:2023alp}. ATLAS quotes $f^{-1}=0.04$--$0.09\TeV^{-1}$ and its event-generator convention is $g_a=4f^{-1}$; matching the $g_a aF\widetilde F/4$ coefficient to Eq.~\eqref{eq:hnllagrangian} gives $g_{a\gamma\gamma}=4f^{-1}=0.16$--$0.36\TeV^{-1}$. Thus the $m_a=300\GeV$ benchmarks are excluded at $g_{a\gamma\gamma}=1\TeV^{-1}$ if the ALP is prompt and has $\mathrm{BR}(a\to\gamma\gamma)=1$, as it does for the stated mass ordering and decay model. This leaves the 50 and 100 GeV HNL grid points not excluded by these cited photon-only collider searches; it does not establish their global viability. The three background samples at each energy contain
$10^6$ events apiece. Because the present signal grid is available at LHE
level, signal and backgrounds are analyzed consistently at parton level; no
detector-level comparison is made in this appendix.

\begin{table}[tbp]
\centering
\caption{Sequential selection for the HNL-mediated
$\gamma\gamma+\met$ analysis. The missing transverse-momentum vector is the
vector sum of final-state neutrino momenta.}
\label{tab:hnlselection}
\small
\begin{tabular}{ll}
\toprule
Stage & Requirement\\
\midrule
Diphoton multiplicity & At least two photons\\
Photon acceptance & $p_T^{\gamma_1}>30\GeV$, $p_T^{\gamma_2}>20\GeV$,
$|\eta_{\gamma_{1,2}}|<2.5$\\
Missing momentum & $\met>0.05\sqrts$\\
Diphoton recoil & $p_T^{\gamma\gamma}>0.05\sqrts$\\
Azimuth and balance & $\Delta\phi(\gamma\gamma,\vec p_T^{\rm miss})>2.8$,
$|p_T^{\gamma\gamma}-\met|/(p_T^{\gamma\gamma}+\met)<0.20$\\
Resonance window & $|m_{\gamma\gamma}-m_a|<0.10m_a$\\
\bottomrule
\end{tabular}
\end{table}

\begin{figure}[tbp]
\centering
\includegraphics[width=.82\textwidth]{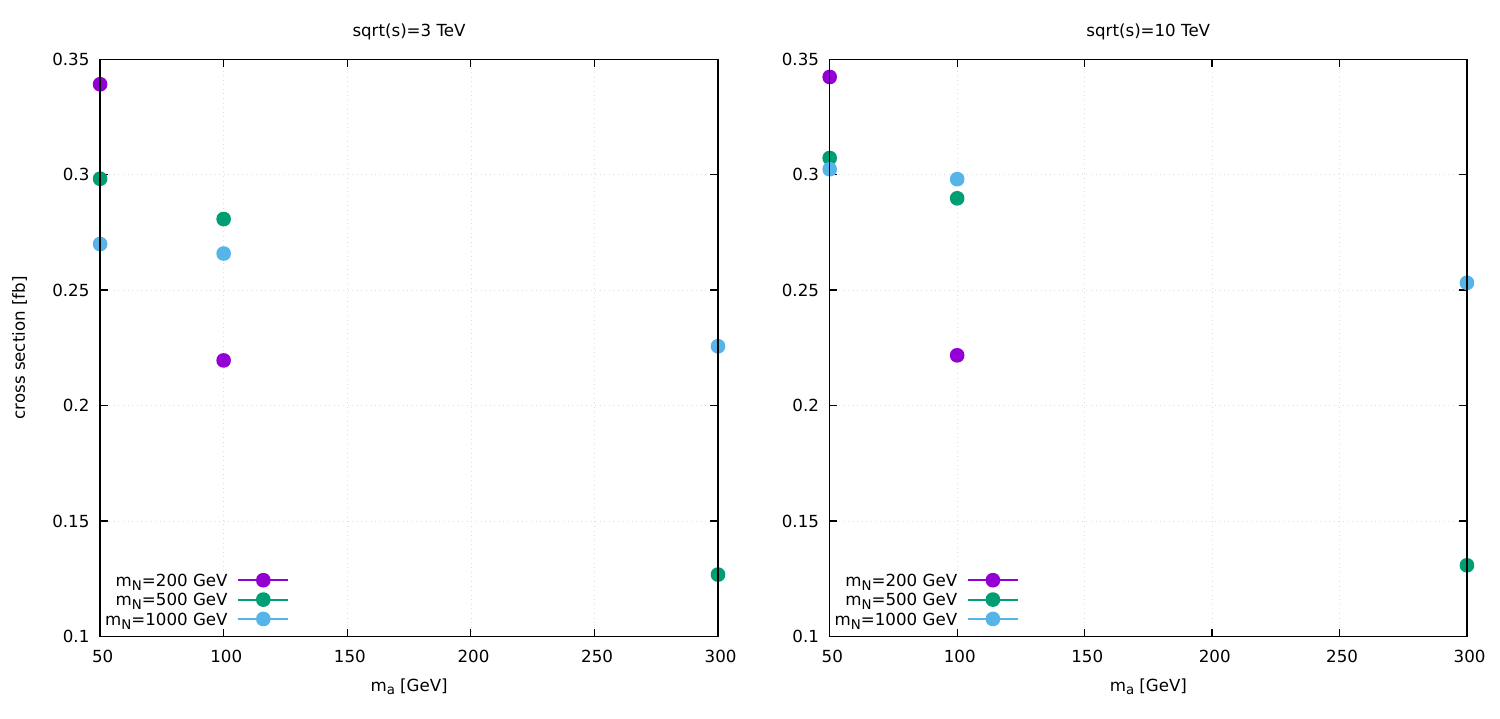}
\caption{Inclusive HNL-mediated signal cross sections at the two collider
energies for the reference couplings. Only kinematically allowed grid points
with $m_N>m_a$ are shown.}
\label{fig:hnlxsec}
\end{figure}

Figure~\ref{fig:hnlxsec} shows rates of $0.1268$--$0.3392$ fb at 3 TeV and
$0.1308$--$0.3423$ fb at 10 TeV. Their mild energy dependence is accompanied
by a strong mass dependence in the selected acceptance, because the fractional
requirements $\met,p_T^{\gamma\gamma}>0.05\sqrts$ become much harder at
10 TeV. The cross sections include the reference branching fraction generated
from Eq.~\eqref{eq:hnlwidths}; they must not be rescaled independently of the
ALP width when $\gagg$ or other ALP decay modes are changed.

\begin{figure}[tbp]
\centering
\includegraphics[width=.84\textwidth]{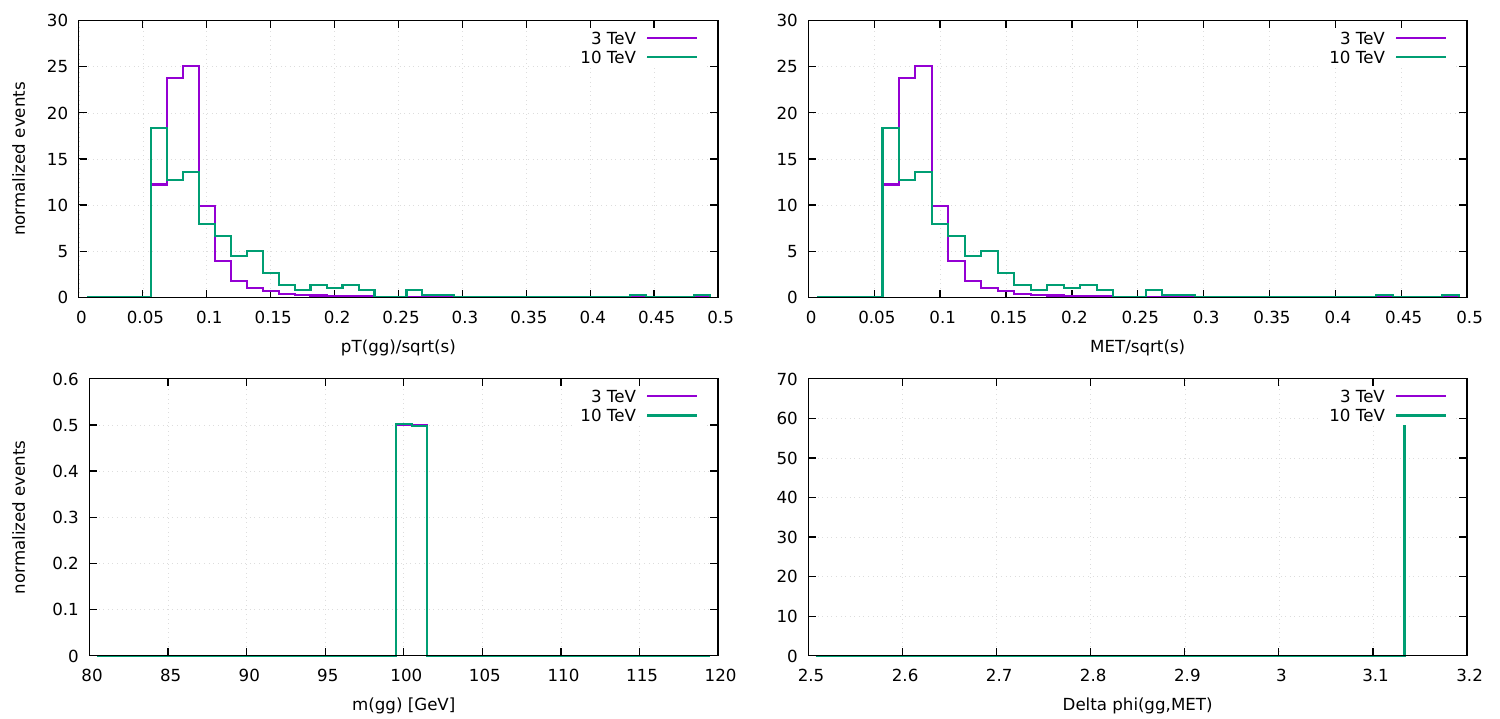}
\caption{Normalized parton-level signal distributions for representative HNL
benchmarks. The narrow $m_{\gamma\gamma}$ peak follows from the on-shell ALP,
while transverse recoil against the invisible system produces the
back-to-back azimuthal topology.}
\label{fig:hnlkinematics}
\end{figure}

\begin{figure}[tbp]
\centering
\includegraphics[width=.94\textwidth]{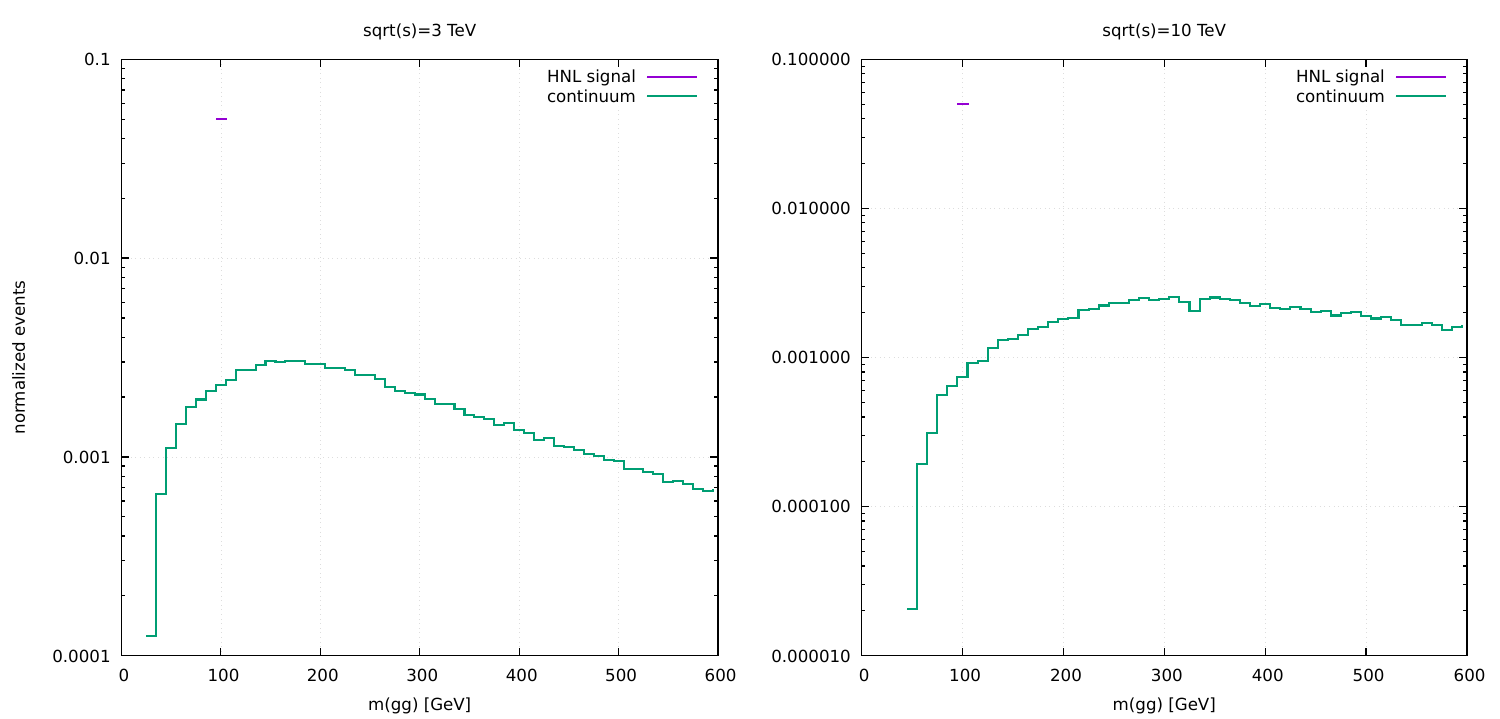}
\caption{Diphoton invariant mass for signal and continuum
$\nu\bar\nu\gamma\gamma$ background at 3 and 10 TeV. Histograms are
normalized to unit area to expose shape discrimination rather than rate.}
\label{fig:hnlmass}
\end{figure}

Figures~\ref{fig:hnlkinematics} and~\ref{fig:hnlmass} explain the selection.
At LHE level the two-body ALP decay gives an artificially sharp mass peak and
nearly exact transverse balance. Consequently, the recoil and balance cuts
remove little signal after the $\met$ requirement, whereas the final mass
window suppresses the smooth continuum by roughly two orders of magnitude.
Photon-energy resolution, showering, isolation, acceptance holes and
beam-induced backgrounds would broaden these idealized structures and must be
included before an experiment-level sensitivity is quoted.

\begin{table}[tbp]
\centering
\caption{Generated HNL-channel background inputs. Each entry contains
$10^6$ unweighted events. The continuum denotes
$\nu\bar\nu\gamma\gamma$; $ZH$ uses $Z\to\nu\bar\nu$ and
$H\to\gamma\gamma$.}
\label{tab:hnlbackgrounds}
\small
\begin{tabular}{rrrr}
\toprule
$\sqrts$ [TeV] & $\sigma(\nu\bar\nu\gamma\gamma)$ [fb] &
$\sigma(ZH)$ [fb] & $\sigma(\gamma\gamma)$ [fb]\\
\midrule
3  & 86.17  & $4.07\times10^{-4}$ & 62.14\\
10 & 103.28 & $3.65\times10^{-5}$ & 5.59\\
\bottomrule
\end{tabular}
\end{table}

\begin{figure}[tbp]
\centering
\includegraphics[width=.84\textwidth]{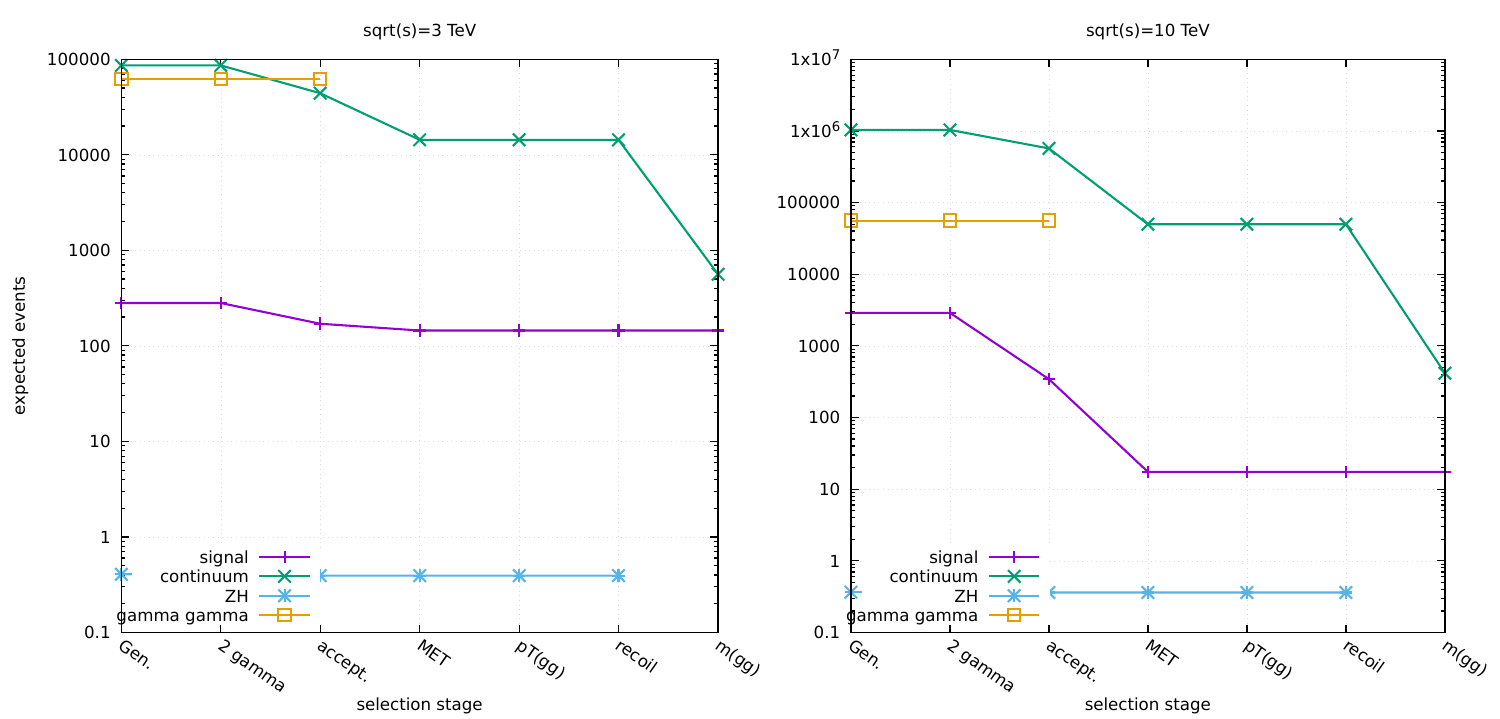}
\caption{Signal and background cut flow for the reference point
$(m_a,m_N)=(100,500)\GeV$, with $1\,\mathrm{ab}^{-1}$ at 3 TeV and
$10\,\mathrm{ab}^{-1}$ at 10 TeV.}
\label{fig:hnlcutflow}
\end{figure}

Figure~\ref{fig:hnlcutflow} shows that photon
acceptance and the scaled missing-momentum threshold control the signal loss.
At 10 TeV the latter reduces the reference signal from 343 to 17.4 events,
which explains why the larger luminosity does not improve every mass point.
The pure diphoton background is eliminated by genuine missing momentum at
parton level; this should not be confused with its detector-level rejection,
where energy mismeasurement can generate a nonzero tail.

\subsection{Yields, significance and indicative coupling reach}

For each mass point we use
\begin{equation}
 Z_{1\%}=\frac{S}{\sqrt{S+B+(0.01B)^2}},
 \label{eq:hnlsignificance}
\end{equation}
matching the cut-based convention used above. A background component with no
surviving Monte Carlo events is assigned a conservative three-event
95\% Poisson upper-count floor multiplied by its event weight. This affects
the final $B$ in table~\ref{tab:hnlresults} of
appendix~\ref{app:hnlresults}, notably for the pure-diphoton
sample, but has little numerical impact where the continuum is large.

\begin{figure}[tbp]
\centering
\includegraphics[width=.84\textwidth]{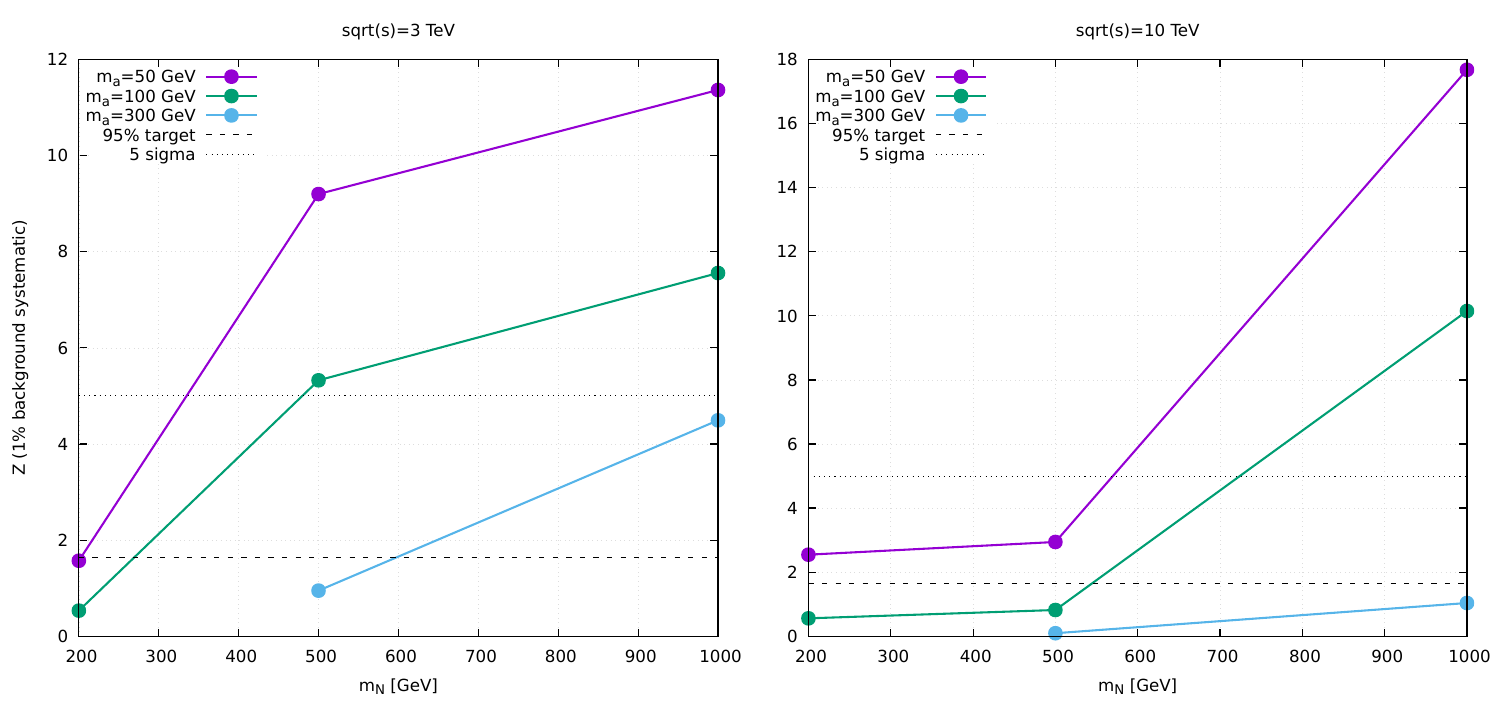}
\caption{Systematics-aware significance versus $m_N$ for each ALP mass at
3 and 10 TeV. Lines only guide the eye between independently generated mass
points.}
\label{fig:hnlsignificance}
\end{figure}

\begin{figure}[tbp]
\centering
\includegraphics[width=.84\textwidth]{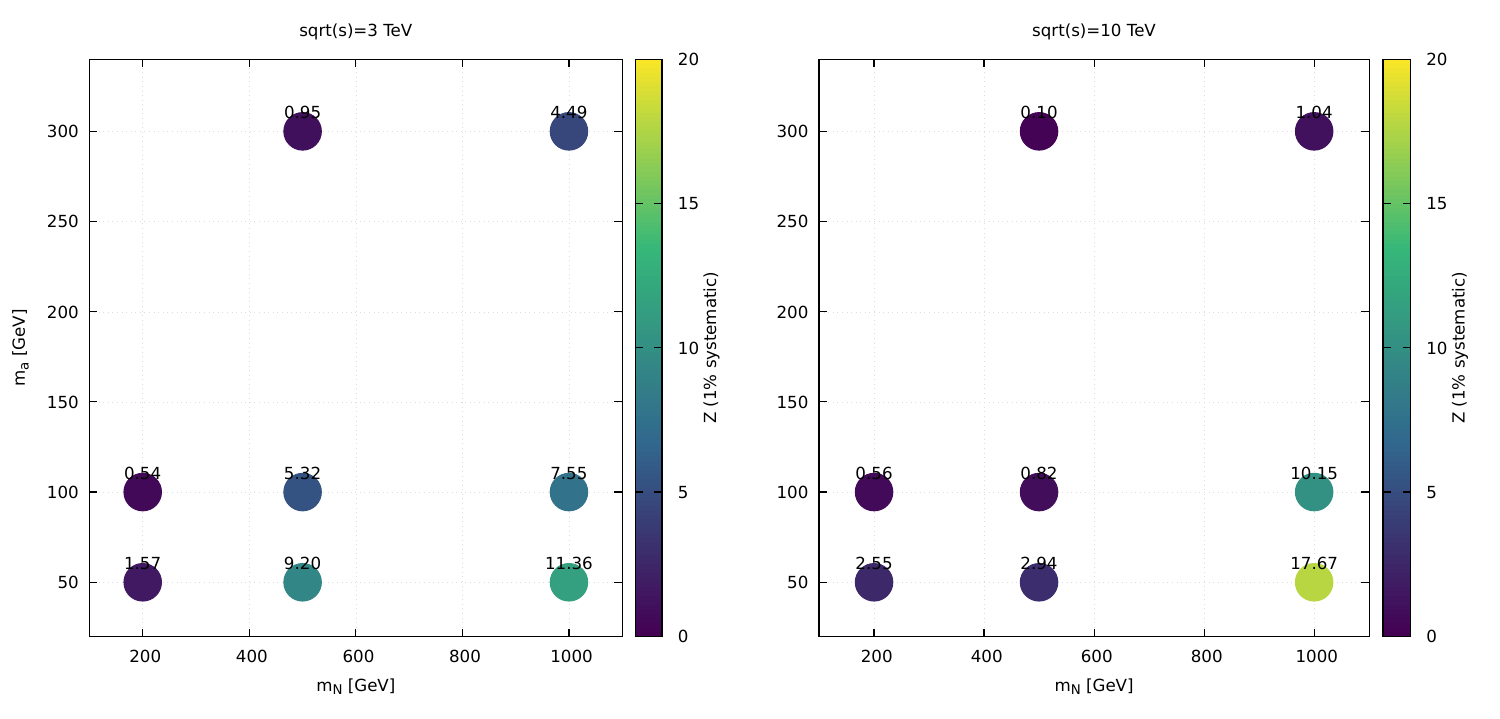}
\caption{Generated mass-plane values of $Z_{1\%}$. No interpolation outside
the simulated nodes, and no result for $m_N\leq m_a$, is implied.}
\label{fig:hnlplane}
\end{figure}

The HNL matrix element is resonantly dominated: a truth-level pairing of each final-state neutrino with the diphoton ALP places 93.4--95.0\% of generated events within five generated $N_1$ widths of $m_N$ over the 16-point grid. These samples therefore describe on-shell $N_1$ production and decay, not merely virtual-HNL exchange.

We retain this appendix to document the resonant topology and identify the detector simulation needed next; it is a parton-level topology and generator diagnostic only. The $m_{\gamma\gamma}$ window is nearly 100\% efficient because the ALP is on shell and photons are unsmeared; it is not a reconstructed mass selection. No showering, photon isolation, cluster merging, calorimeter energy resolution, beam-induced background, or fake/merged-photon model is included. For a boosted two-body decay, $\Delta R_{\gamma\gamma}\sim2m_a/p_T^a$, so separations of 0.05--0.1 occur for $p_T^a\sim1$--2 TeV at $m_a=50\GeV$. In our 10 TeV samples, the generated median $\Delta R_{\gamma\gamma}$ is 0.31, 0.60, and 1.36 for $m_N=1000,500,200\GeV$, respectively; about 0.1\% of events have $\Delta R<0.1$. These numbers do not establish photon reconstruction efficiency. The $Z_{1\%}$ values in table~\ref{tab:hnlresults} are parton-level bookkeeping quantities, not experimental sensitivities, exclusions, or coupling reaches. We omit a fixed-acceptance one-parameter coupling rescaling because varying $c_N$, $V_{\mu N}$, or $f_N$ changes production amplitudes and widths/branching fractions; a physical bound requires regeneration in a defined gauge-consistent model.

\section{Complete HNL diphoton results}
\label{app:hnlresults}

Table~\ref{tab:hnlresults} lists cross sections, parton-level selection efficiencies, yields, and bookkeeping significances for the HNL feasibility samples.

\footnotesize
\begin{longtable}{rrrrrrrr}
\caption{Parton-level HNL feasibility results. The reference parameters are $|V_{\mu N}|=10^{-2}$, $c_N=1$, $f_N=1\TeV$, and $g_{a\gamma\gamma}=1\TeV^{-1}$. $Z_{1\%}$ is not a detector-level sensitivity or physical coupling bound.}\label{tab:hnlresults}\\
\toprule
$\sqrts$ [TeV] & $m_a$ [GeV] & $m_N$ [GeV] & $\sigma$ [fb] & $\epsilon$ &
$S$ & $B$ & $Z_{1\%}$\\
\midrule
\endfirsthead
\toprule
$\sqrts$ [TeV] & $m_a$ [GeV] & $m_N$ [GeV] & $\sigma$ [fb] & $\epsilon$ &
$S$ & $B$ & $Z_{1\%}$\\
\midrule
\endhead
3&50&200&0.339&0.0614&20.8&152.2&1.57\\
3&50&500&0.298&0.551&164.2&152.2&9.20\\
3&50&1000&0.270&0.814&219.8&152.2&11.36\\
3&100&200&0.220&0.0601&13.2&562.1&0.54\\
3&100&500&0.281&0.515&144.6&562.1&5.32\\
3&100&1000&0.266&0.808&214.8&562.1&7.56\\
3&300&500&0.127&0.305&38.7&1415.0&0.95\\
3&300&1000&0.226&0.846&191.0&1415.0&4.49\\
10&50&200&0.342&0.00486&16.6&26.0&2.55\\
10&50&500&0.307&0.00650&20.0&26.0&2.94\\
10&50&1000&0.302&0.111&336.5&26.0&17.67\\
10&100&200&0.222&0.00536&11.9&415.4&0.56\\
10&100&500&0.290&0.00602&17.4&415.4&0.82\\
10&100&1000&0.298&0.0902&268.8&415.4&10.15\\
10&300&500&0.131&0.00526&6.9&3610.9&0.10\\
10&300&1000&0.253&0.0290&73.5&3610.9&1.04\\
\bottomrule
\end{longtable}

\section{Source files and reproducibility}
\label{app:reproducibility}

The event generation used MadGraph5\_aMC@NLO 3.6.2 and Pythia 8.311; the
analysis environment uses ROOT 6.28/10 and the Delphes installation supplied
with the local MadGraph tool chain. The cut-based figures and tables are stored under
\path{alp_neutrino_manuscript_results/}; the classifier figures and tables
are stored under
\path{alp_neutrino_tmva_final_manuscript_results/}. The complete
machine-readable values remain in the corresponding CSV files. The
supplementary release archive contains these tables and plot inputs, together
with the publication figures and analysis scripts. This document rounds values
for typesetting but does not replace those CSV files for numerical reuse.

The benchmark definitions, UFO validation and run provenance are in
\path{ALP_NEUTRINO_BENCHMARKS.md}, \path{MODEL_ISSUES_RESOLUTION.md} and the
generated sample manifests. The Grid-2 numbers in
table~\ref{tab:modelcomparison} are independently recoverable from
\path{cross_section_comparison/diphoton_signal_cross_sections.csv}. The compact
release archive does not include all event-level samples or ROOT/TMVA trees;
these remain available from the corresponding author on reasonable request.

\end{document}